\documentclass[10pt,journal,compsoc]{IEEEtran}

\usepackage{tabularx}
\usepackage{multirow}
\usepackage{hyperref}
\usepackage{paralist}
\usepackage{graphicx}
\usepackage[autolanguage]{numprint}
\usepackage{xcolor}
\usepackage{mdframed,lipsum}
\usepackage{diagbox}
\usepackage{tabularx}
\usepackage{balance}
\usepackage[inline]{enumitem}
\usepackage{listings,setspace,textcomp}
\usepackage{newfloat,caption}
\usepackage{subcaption}
\usepackage{makecell}
\usepackage{pdfpages}

\usepackage{amssymb}
\usepackage{pifont}
\usepackage[justification=centering, font=small]{caption}

\newcommand{\TSErevision}[1]{\textcolor{black}{#1}}

\newcommand{\xmark}{\ding{53}}%

\definecolor{bluekeywords}{rgb}{0.13,0.13,1}
\definecolor{greencomments}{rgb}{0,0.55,0.2}
\definecolor{redstrings}{rgb}{0.9,0,0}

\DeclareFloatingEnvironment[fileext=frm,placement={!ht},name=Listing]{listing}

\begin{document}
%
\title{Automated Classification of Overfitting Patches with Statically Extracted Code Features}

\author{He~Ye,
        Jian Gu,
        Matias~Martinez,
        Thomas Durieux,
        and~Martin~Monperrus
        
 \IEEEcompsocitemizethanks{\IEEEcompsocthanksitem He Ye, Jian Gu,  Thomas Durieux, and Martin Monperrus are with the KTH Royal Institute of Technology, Sweden. E-mails: heye@kth.se, jiagu@kth.se,  thomas@durieux.me, and martin.monperrus@csc.kth.se \protect\\

 \IEEEcompsocthanksitem Matias Martinez is with Université Polytechnique Hauts-de-France, France. E-mail: matias.martinez@uphf.fr }%
 \thanks{}
}

\markboth{}%
{}

\IEEEtitleabstractindextext{%
\begin{abstract}
Automatic program repair (APR) aims to reduce the cost of manually fixing software defects. However, APR suffers from generating a multitude of overfitting patches, those patches that fail to correctly  repair the defect beyond making the tests pass.
This paper presents a novel overfitting patch detection system called ODS to assess the correctness of APR patches.
ODS first statically compares a patched program and a buggy program in order to extract code features at the abstract syntax tree (AST) level, for the single programming language Java. 
Then, ODS uses supervised learning with the captured code features and patch correctness labels to automatically learn a probabilistic model. 
The learned ODS model can then finally be applied to classify  new and unseen program repair patches. 
We conduct a large-scale experiment to evaluate the effectiveness of ODS on patch correctness classification based on \numprint{10302} patches from Defects4J, Bugs.jar and Bears benchmarks. 
The empirical evaluation shows that ODS is able to correctly classify 71.9\% of program repair patches from 26 projects, which improves the state-of-the-art. 
ODS is applicable in practice and can be employed as a post-processing procedure to classify the patches generated by different APR systems.
\end{abstract}

\begin{IEEEkeywords}
Automatic program repair; Patch assessment; Overfitting patch;  Code features
\end{IEEEkeywords}}

\maketitle

\IEEEdisplaynontitleabstractindextext

\IEEEpeerreviewmaketitle


\section{Introduction}\label{sec:introduction}

\begin{figure*}[th]
  \centering
\includegraphics[width=0.95\textwidth]{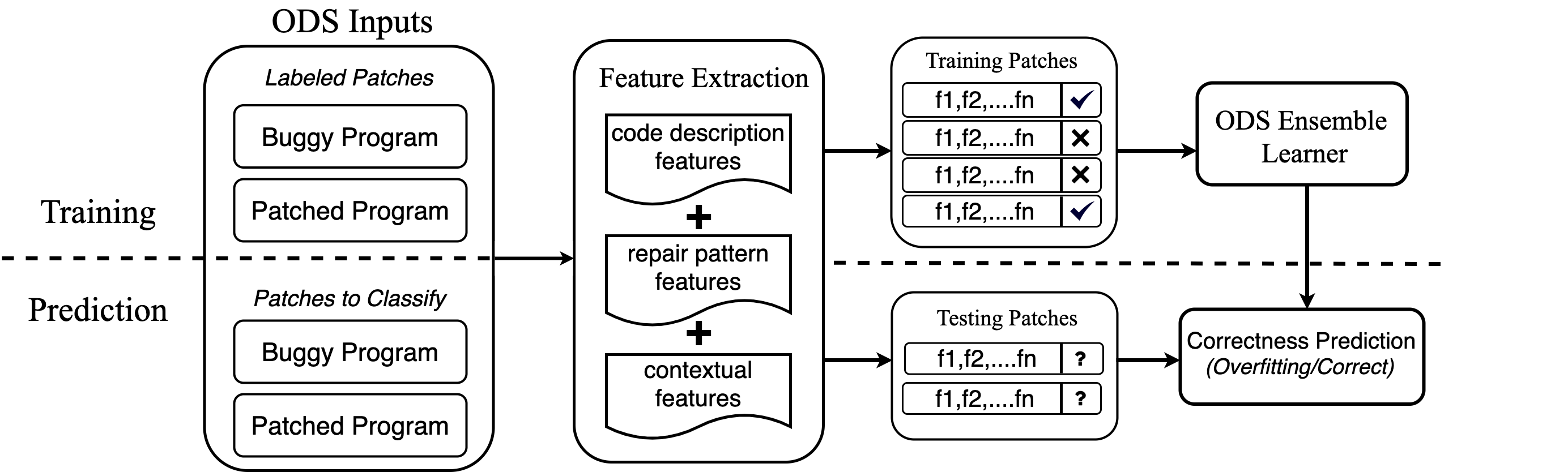}
 \caption{Overall process of ODS to detect overfitting patches}
\label{fig:ods}
\end{figure*}

\IEEEPARstart{P}{rogram} repair consists of automatically generating patches for defects \cite{Monperrus2015,ahbik-automated-repair-CACM}. 
In test-suite based repair, a test suite acts as an executable specification to drive the patch search.
A significant challenge for test-suite based repair is that it is possible to generate a patch that makes the whole test suite pass, yet that is still incorrect: this is known as \textit{the overfitting patch problem} \cite{cure-worse-15,kali,Le:overfitting,ISSTA19-bytecode-repair}.

The overfitting patch problem results in low value generated program repair patches, which significantly affect the program repair applicability in practice. 
To overcome this problem, research \cite{issta17-difftgen,le:reliability-patch-assess,drr} employs automatic test generation techniques to construct oracles, based on the human-written patches. All those techniques need a human-written patch to construct the oracles. They fit well scientific evaluations of program repair, but such a ground truth patch is usually not available in practice.  
\TSErevision{
Notably, Xiong et al. \cite{patchsim}'s technique is the first one that does not require a human-written patch as a ground truth. 
It works by measuring the dynamic behavior changes before and after patching the defect, based on the execution traces of test cases.   
We refer to their technique as PatchSim in our paper. 
}
The main shortcoming of PatchSim is that it is time-consuming because it requires compilation, code instrumentation and execution of several tests.

To solve this problem, recent works have proposed static approaches for overfitting patch detection.
Tian et al. \cite{ase20Bert} leverage BERT
\cite{Bert} to automatically extract code features for overfitting patch detection, and Wang et al. \cite{ASE20Wang} employ eight static features for classifying overfitting patches.
This shows that using code features with machine learning is a promising technique for patch assessment and this is the domain where our paper makes a novel contribution.

In this paper, we present ODS, a novel system to statically classify overfitting patches. ODS is an acronym for \textbf{O}verfitting \textbf{D}etection \textbf{S}ystem.
ODS is founded on the hypothesis that overfitting patch detection can be made across program repair systems and across software projects, because there exist code features that capture universal correctness properties for classifying overfitting patches. 
ODS works as follows: it extracts \textit{202} code features per patch, that are statically extracted at the abstract syntax tree (AST) level. 
ODS features include:
1) \textit{150} code description features representing characteristics of patch ingredients (e.g., variables, operators, statements and AST operators, etc) at different granularity in the modified source code and code surrounding them. For example, we capture whether a patch uses a local variable or a global variable (\textit{localVar} or \textit{globalVar});
2) \textit{26} repair pattern features encoded with human knowledge on repair strategies. For example, we capture whether a patch moves existing code to a different location (\textit{codeMove});
3) \textit{26} contextual syntactic features that encode the scope and similarity information in the source code. \TSErevision{For example, we capture a feature that encodes whether there exists a variable of the same type in the surrounding context (\textit{similarObjectType}).}
Then, ODS employs an ensemble learning model based on decision trees to produce the probability of a patch being overfitting.  

Our training set contains patches for the bugs from Defects4J~\cite{defects4j}, Bugs.jar~\cite{bugjar} and Bears~\cite{Bears2019}, generated by the work of Durieux et al. \cite{repairthemall}. 
Since the correctness of \numprint{62824} patches from \cite{repairthemall} were unknown, we set up a systematic methodology to build a sound labeled training set of patches:
1) sanity checks to discard duplicated patches (\numprint{37777}) and  non-plausible patches (\numprint{9190});
2) automatic patch assessment with generated tests by Evosuite \cite{evosuite} to obtain \numprint{8299} overfitting patches;
3) a collection of \numprint{2003} human-written patches.
Overall, we obtain a total of \numprint{10302} labeled patches, which enables us to perform a comprehensive evaluation of patch assessment. 

On this data, we show that ODS achieves a precision of 94.7\% and a recall of 70.0\% for overfitting patch detection.
Those experiments show that ODS outperforms PatchSim, BERTFeature \cite{ase15-tests} and patch assessment technique based on eight static code features \cite{ASE20Wang}. 
Beyond effectiveness, ODS is fast. Since ODS is static and does not require compilation and execution, it is faster than the dynamic related work: in our experiment, ODS is 138 times faster than PatchSim.
This is arguably an important factor for practitioners to use overfitting patch detection technique in practice.

\TSErevision{Our work is novel along the following lines: 
\begin{enumerate*}
\item Our code feature set is novel: it contains hundreds of new features, of a more diverse nature (202 features versus 35 features in \cite{prophet} and 8 in \cite{ASE20Wang});
\item Our paper is the first  to analyze the complementarity between dynamic patch assessment and static patch assessment;
\item We are the first to measure and report on the correct patch ratio with and without overfitting detection over 19 repair tools, showing that  ODS would save developer time by avoiding presenting them overfitting patches for 18/19 tools;
\item We present unique and original results on project independence for overfitting patch classification (the fact that the overfitting knowledge gained on one project can be applied to another project);
\item Our dataset is novel: we provide a curated dataset of \numprint{10302} labeled patches to the research community, which is 10 times larger than the dataset of Wang et al. \cite{ASE20Wang} with 902 patches.
\end{enumerate*}
}

To sum up, our contributions are: \begin{itemize}
\item A novel technique called ODS for identifying overfitting patches generated by program repair tools. ODS is based on original code features that are statically extracted from the AST of the buggy program and the patched program. ODS defines and implements a novel set of \textit{202} code features that syntactically and semantically represent a program repair patch.

\item  A large-scale experiment showing that: 
1) ODS achieves higher effectiveness than the state-of-the-art in terms of accuracy;
2) Overfitting patch detection can be made project independent: the overfitting knowledge learned on one software project can be applied on another one.

\item A dataset of \textit{\numprint{10302}} labeled correct and overfitting patches from \textit{\numprint{62824}} unlabeled program repair patches from \cite{repairthemall}. We consider this dataset as a valuable asset for future research on overfitting patch assessment.

\item We fully implemented ODS as a publicly-available open source tool for overfitting patch detection~\cite{ourrepo}. ODS comprises a powerful engine for source code feature extraction in Java programs, and an optimized machine learning model that can be embedded in future program repair tools.  
\end{itemize}

\section{The ODS Approach for Overfitting Detection}

We present ODS, a novel automatic patch assessment technique for program repair systems.

\subsection{Overview of ODS}

\autoref{fig:ods} gives the overall process of the ODS approach, it consists of two phases: the training phase (shown in the top half) and the prediction phase (shown in the bottom half).
In the training phase, ODS learns a probability model from labeled training patches whose correctness is known.
In the prediction phase, the learned ODS model is used to predict the correctness of new and unseen patches whose correctness is unknown. 

\emph{Training phase.}  ODS takes as inputs a set of buggy programs and their corresponding patched programs, each of which has a correctness label: \textit{overfitting} or \textit{correct}. 
ODS performs static analysis to extract the AST's differences between the buggy and patched programs, those differences are encoded as \emph{feature vectors} \cite{learningToBlame}. 
Feature vectors refer to real numbers in an n-dimensional space, where each column describes a particular aspect of the input (see the bottom of  \autoref{fig:odsexample}).  
For each pair of program source code, ODS extracts three categories of code features to form feature vectors (see \autoref{sec:featurs}).
Then, ODS performs supervised learning on a training set of feature vectors with labels.
ODS learns a probability model that can be used to predict testing patches with a confidence score.

\emph{Prediction phase.} 
For a new and unseen patch, the code features are first extracted in the same manner conducted in the training patches.
Then, the learned ODS model is fed with the extracted code features. 
For each patch, the ODS outputs a probability score that represents the likelihood of its correctness.

\subsection{Feature Extraction}
\label{sec:featurs}

\begin{table*}
\footnotesize
\renewcommand{\arraystretch}{1.6}
\begin{tabularx}{0.97\textwidth}{@{}lcXXc@{}}
\hline
Feature Category & Type &Description & Feature Examples& \# Total \\

\hline

Operators&binary& whether patch contains binary arithmetic operators, unary arithmetic operators, relational operators or bitwise operators  &opAdd, opDiv, opEqual, opGreaterEqual, opLessThan,opMod,uopDec, uopInc,  etc& 14 \\

\hline

Variables & binary &  whether patch variables are local, global, abstract, primitive, enumeration & localVar, globalVar, abstVar, primVar, enum & 5\\

\hline

Statements & binary & whether patch occurs in assignment, condition, loop, try-catch, function invocation, return, branch, constance, &  assignConst, assignLhs, assignZero, callee, callArgument, StmtCond, StmtCall, stmtLoop, memberAccess, funcArgument, etc & 21\\

\hline
AST operations &binary &  whether AST actions are  insertion/removals of operators, variables or statements & insertStmt, replaceCond, replaceStmt, removePartialIf, removeWholeBlock &10\\
\hline

\multicolumn{4}{c}{Total } &50\\
\hline
\end{tabularx}
\caption{ODS code description features }
\label{tab:codeDescriptionFeatures}
\end{table*}

Following Long and Richard \cite{prophet}, we model a program as a set of feature vectors based on human knowledge. 
We analyze the source code files that have changed between the buggy and the patched version.
The difference between those programs is known as a \emph{patch}.
Given a pair of buggy and patched source code files, 
we first compute the AST of each source code file affected by the patch, then we execute an AST differencing algorithm over those ASTs. This computes a list of AST operations (aka \emph{edit-script}) that captures the transformations between one AST (which represents the buggy file) into the other (which represents the patched file). 
We call this list an `AST diff'.

Then, we transform the AST diff into fixed-length feature vectors.
For multiple AST diffs from a patch that affects two or more files (which means the patch modifies different files to repair a bug), we accumulate the occurrence per feature vector from each AST diff.
As a result, we obtain a set of fixed-length feature vectors that represents a patch that fixes a buggy program.

\subsubsection{Feature Categories}
\label{sec:ods_features}

We group the ODS features into three categories: code description features, repair pattern features, and contextual syntactic features.

\textbf{Code Description Features.}
These features describe the characteristics of code elements at different granularities.
In \autoref{tab:codeDescriptionFeatures}, we present our 50 code description features of four categories: operators, variables, statements and AST operations. 
The first column gives the category, and the second column indicates the type of each feature category. Here, all code description features are binary features.
We describe them in the third column and show examples of feature names in the fourth column. 
The last column summarizes the total number of extracted features in each category.
We now describe code description features per category.

\emph
{Operators:} ODS extracts operator features in four different types if they are used in a patch: binary arithmetic, unary arithmetic, relational and bitwise operators.  
\emph
{Variables:} 
ODS extracts the characteristics of variables involved in the patch such as scope (e.g., local or global) and type  (e.g., primitive).
\emph
{Statement:} 
ODS captures the types of the statements affected by a patch (e.g., control flow, assignment, invocation, and return, etc).
\emph
{AST operations:} 
ODS creates AST operation features by combining:
\begin{inparaenum}[\it a)]
\item the type of operation from the AST diff (4 AST-level operations: UPD, ADD, DEL and MOV); and 
\item  code elements affected by the AST-level operation.
For example, feature \textit{insertStmt} indicates the patched program inserts a new statement to fix the bug.
\end{inparaenum}

We also extract features from statements that surround the patch.
In particular, we consider features from:
\begin{inparaenum}[a)]
\item the modified statements (we call them SRC features);
\item at most three statements in the same block before the modified statements (\textit{FORMER} features); and
\item at most three statements in the same block after the modified statements (\textit{LATTER} features).
\end{inparaenum}
This means that ODS considers, for each patch, 150 code description features (i.e., 50 features from \textit{SRC} + 50 features from \textit{FORMER} + 50 features from \textit{LATTER}).

\textbf{Repair Pattern Features.} 
To help ODS learn the semantic information of a patch, we include repair patterns extracted by human experts.   
We use the taxonomy of repair patterns by Sobreira et al. \cite{defects4J-dissection}, which defines a list of repair patterns and actions, based on manual analysis of 395 Defects4J human-written patches. Madeiral et al. \cite{Madeiral2018} have shown that those patterns can be automatically extracted, and we reuse the corresponding tool. 
ODS integrates 26 repair patterns from \cite{Madeiral2018}.
\autoref{tab:repairPatternFeatures} presents the 26 repair patterns considered by ODS. 
The first column contains the categories of the repair patterns. The type of repair patterns is given in the second column.
All repair patterns are binary features. 
We present the detailed repair patterns in the third column and summarize the total number by category in the last column. 

\emph{Wraps-with patterns} include patterns that wrap or unwrap buggy statements with a conditional branch, try-catch block, or loop.
For example, \emph{wrapsIf} indicates a potential buggy statement is wrapped with a conditional logic using an \emph{if} expression.
\emph{Expression patterns}  occur in patches that modify existing logic or arithmetic expressions, return expressions or assignments of boolean variables.
\emph{Conditional Patterns} include patterns that add or remove conditional blocks.
\emph{Null check patterns} are related to the
addition of conditional expressions or expansion of existing
ones with null-checks, including positive null checks (check for nullity) and negative null checks (check for non-nullity).
\emph{Other patterns} include other useful patterns such as repeating the same change in different locations (e.g., \textit{copyPaste}); move the existing code to different locations (e.g., \textit{codeMove}); removal or replacement of a single line (e.g., \textit{singleLine}), etc.

\begin{table}
\footnotesize
\renewcommand{\arraystretch}{1}

\begin{tabularx}{0.47\textwidth}{@{}llXc@{}}
\hline
Category  &Type & Repair Patterns & \#Total \\

\hline

\multirow{6}{*}{ Wraps-with }& \multirow{6}{*}{binary }& wrapsIf& \multirow{6}{*}{ 9}\\
&&wrapsElse  \\
&& wrapsLoop\\
&&wrapsTryCatch \\
&&unwrapTryCatch\\ 
&&wrapsIfElse\\
&&unwrapIfElse\\
&&wrapsMethod\\
&&unwrapMethod \\
\hline

\multirow{4}{*}{ Expression }&\multirow{4}{*}{ binary}& expLogicExpand,& \multirow{4}{*}{4} \\
&&expArithMod,\\
&& expLogicReduce,\\ 
&&expLogicMod\\

\hline
\multirow{4}{*}{Conditional} &\multirow{4}{*}{ binary}& condBlockOthersAdd & \multirow{4}{*}{4} \\
&&condBlockRem,\\
&& condBlockExcAdd,\\ 
&&condBlockRetAdd\\
\hline
\multirow{2}{*}{Null checks }&\multirow{2}{*}{ binary}&missNullCheckP &\multirow{2}{*}{ 2}\\
&&missNullCheckN\\
\hline
\multirow{7}{*}{Other patterns} & \multirow{7}{*}{binary}&codeMove & \multirow{7}{*}{7} \\
&&copyPaste \\
&&wrongVarRef\\ 
&&wrongMethodRef\\
&&singleLine\\ 
&&constChange \\ 
&&notClassified \\
\hline
\multicolumn{3}{c}{Total} &26\\
\hline
\end{tabularx}
\caption{ODS repair pattern features }	
\label{tab:repairPatternFeatures}
\end{table}

\begin{table}
\footnotesize
\renewcommand{\arraystretch}{1.1}
	
\begin{tabularx}{0.48\textwidth}{@{}llXc@{}}
\hline
Category & Type & Contextual Features & \#Total \\
\hline

\multirow{7}{*}{Type}& \multirow{7}{*}{string} & typeOfFaultyStatementAfter1 & \multirow{7}{*}{9}\\
&&typeOfFaultyStatementAfter2\\
&&typeOfFaultyStatementAfter3,\\
&&typeOfFaultyStatementBefore1\\
&&typeOfFaultyStatementBefore2\\
&&typeOfFaultyStatementBefore3,\\
&&typeOfFaultyStatement\\
&&typeOfFaultyStatementParent\\
&&faultyClassExceptionType \\

\hline

\multirow{6}{*}{Method} & \multirow{6}{*}{binary} & methodCallWithNullGuard & \multirow{6}{*}{7}\\
&&methodCallWithTryCatch,\\
&&inSynchronizedMethod,\\ 
&&hasObjectiveMethodCall,\\
&&methodThrowsException,\\
&&methodCallWithNormalGuard,\\
&&hasInvocationsProneException\\

\hline
\multirow{4}{*}{Similarity}& \multirow{4}{*}{binary} &similarObjectTypeWithNormalGuard & \multirow{4}{*}{4}\\
&&similarObjectTypeWithNullGuard,\\
&&similarPrimitiveTypeWithNormalGuard\\
&&similarPrimitiveTypeWithNullGuard\\
\hline
\multirow{5}{*}{Usage}&\multirow{5}{*}{binary}&fieldNotAssigned & \multirow{5}{*}{6}\\
&&fieldNotUsed,\\
&&localVarNotAssigned\\
&&localVarNotUsed, \\
&&objectUsedInAssignment\\
&&primitiveUsedInAssignment,\\
\hline
\multicolumn{3}{c}{Total} &26\\
\hline

\end{tabularx}
\caption{ODS contextual syntactic features }
\label{tab:contextfeatures}
\end{table}

\textbf{Contextual Syntactic Features.}
The context of a patch can be critical for correctly predicting patch correctness. 
Inspired by \cite{zhongxing-s4r,learningToBlame}, we include contextual features to describe the scope, parent and children’s similarities of modified statements.
In \autoref{tab:contextfeatures}, we present 26 contextual syntactic features considered in ODS.
The first column lists the feature categories, and the second
column gives the type of each feature, namely binary and string. The detailed contextual features are presented in the third column and the numbers are given in the last column. We now describe the categories of such features.

\emph{Type-related} features focus on the type of the modified statements and the type of its surrounded code and parent code. The type features are captured as a string. 
For example, the type of a faulty statement (\emph{typeOfFaultyStatement}) could be an \emph{assignment} and its parent statement type \emph{typeOfFaultyStatementParent} could be a \emph{method} or a \emph{class}. 
\emph{Method} describes the context related to the method invocation, whether the method call happens within a try-catch block.
\emph{Similarity} describes whether there exists object or method calls with the same type (e.g., object type or primitive type) in the scope that is similar.
\emph{Usage} features focus on whether there exists any field or variable never used or assigned in the class.

\begin{figure*}[t]
  \centering
\includegraphics[width=0.98\textwidth, height=10cm]{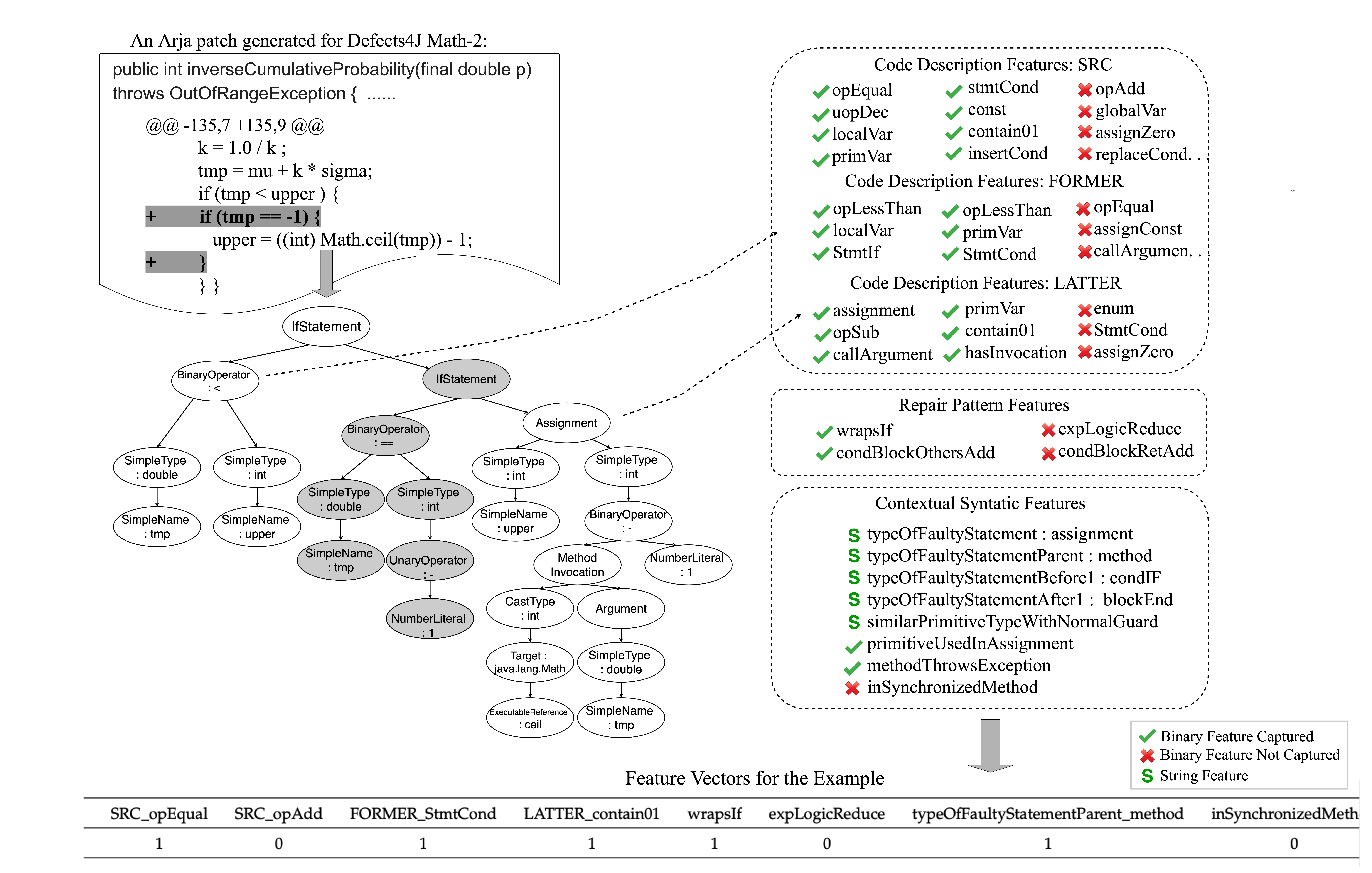}
\caption{An example of ODS feature extraction.}
\label{fig:odsexample}
\end{figure*}

\subsubsection{Feature Vectors}
In total, a patch is described by 202 ODS features:
150 code description features, 26 repair pattern features and 26 contextual syntactic features.
We convert the binary features to an integer to represent feature occurrence, where \textit{False} is converted to \emph{0}, and \textit{True} is converted to \emph{1}. 
We leverage one-hot encoding to transform  string type features into binary indicator features~\cite{dummyVariables}: each string feature with \textit{m} possible values is encoded by a sparse matrix of \textit{m} binary features, with only one active (the active feature is 1, otherwise 0). 
For multiple AST diffs from one patch, we aggregate the corresponding values per feature from each AST diff.

\subsubsection{Example}
\autoref{fig:odsexample} shows an example of how ODS extracts features from a patch generated by Arja \cite{Yuan2017ARJAAR}. 
This patch adds an if-conditional check to wrap an assignment of \textit{upper}. We use the  \checkmark and \xmark~  to indicate the captured and non-captured  binary features respectively, and the symbol $S$ identifies string features.
ODS computes the AST-level changes between the AST from the buggy program and the patched program. The AST diff is presented in gray.
From the AST nodes affected by those changes, ODS starts extracting the features.

ODS extracts the aforementioned three categories of syntactic and semantic features to represent those AST changes (i.e., patch).
For this patch, ODS extracts code description features: for example, features \textit{opEqual} (binary operator '=='), \textit{uopDec} (unary operator '-'), \textit{localVar} (local variable \textit{tmp}) are captured in the patch (i.e., their values receive a \checkmark). On the contrary, other features (e.g., \texttt{assingZero} and \texttt{opAdd}) are not captured (thus their values receive a \xmark).

Similarly,  ODS also extracts the code description features from the former and latter statements around the changed AST (mapped with long-dotted lines). 
ODS identifies 2 out of 26 repair patterns for this example: \textit{wrapsIf} and \textit{condBlockOthersAdd}.
Note that other patterns could not be captured in the patch.
For example, \textit{condBlockRetAdd} is not present because, beyond adding an if condition, the patch does not add a return statement.

ODS captures contextual features, which types can be a string or a binary.
For example, the value of the feature \textit{typeOfFaultyStatementParent} for this patch is a \textit{method}. 
Eventually, ODS converts all binary and string features to integer vectors.
As shown in the bottom of \autoref{fig:odsexample}. 
Notably, ODS converts the string value feature \textit{typeOfFaultyStatementParent} into a binary indicator feature  \textit{typeOfFaultyStatementParent\_method}.

\subsection{Learning Algorithm}
For the classifier, we consider the gradient boosting algorithm \cite{GradientBoosting}.
Gradient boosting is a machine learning technique for regression and classification problems, which produces a prediction model in the form of an ensemble of weak prediction models, typically decision trees \cite{decisiontree}. We choose gradient boosting for the following three reasons. First, gradient boosting has been shown to give state-of-the-art results on many standard classification benchmarks \cite{GBEffective}.  
Second, gradient boosting as an ensemble method helps to reduce variance and bias, and to build a more generalized model. 
Third, the high degree of flexibility that is offered by gradient boosting allowing optimization of different loss functions that can be adjusted to classification models.

\subsection{Implementation}
We fully implemented our approach in the context of Java. 
We implemented ODS feature extraction engine on the top of two libraries: Coming \cite{coming}, which extracts the code description features and contextual syntactic  features, and ADD \cite{Madeiral2018}, which extracts repair pattern features. 
Both tools internally use GumTree
~\cite{Falleri:2014:FAS} to compute AST diffs.
For the ODS classifier, we use off-the-shelf ensemble learning library XGBoost  \cite{xgboost} which is based on decision trees.



\section{Experimental Protocol}
In this section, we present our protocol to evaluate the effectiveness of ODS for classifying overfitting patches. The protocol is designed to answer the following research questions:

\begin{itemize}

\item RQ1 (Effectiveness): To what extent is ODS good at predicting overfitting patches compared with the state-of-the-art?
\item \TSErevision{RQ2 (Complementary analysis): To what extent is ODS complementary to the dynamic patch assessment technique PatchSim?}
\item RQ3 (Project independence): To what extent can ODS be used for overfitting detection on new projects?
\item RQ4 (Feature analysis): Which features are contributing the most to ODS accuracy? 
\end{itemize}

\subsection{Patch Datasets with Overfitting Labels}
\begin{table}[t!]
\scriptsize
\renewcommand{\arraystretch}{1.5}
\begin{tabular}{p{0.25\linewidth}p{0.15\linewidth}p{0.15\linewidth}p{0.15\linewidth}}
\hline
tools & \#overfitting & \#correct & \#total \\
\hline
Arja&52&5&57\\
\hline
ACS&7&15&22\\
\hline
AVATAR&38&19&57\\
\hline
CapGen&41&25&66\\
\hline
Cardumen&10&2&12\\
\hline
DynaMoth&21&1&22\\
\hline
FixMiner&20&12&32\\
\hline
GenProg&45&5&50\\
\hline
Jaid&41&40&81\\
\hline
jMutRepair&17&5&22\\
\hline
Kali&82&6&88\\
\hline
kPAR&52&10&62\\
\hline
Nopol&30&1&31\\
\hline
RSRepair&39&2&41\\
\hline
SequenceR&56&17&73\\
\hline
SimFix&45&22&67\\
\hline
SketchFix&9&16&25\\
\hline
SOFix&2&21&23\\
\hline
TBar&47&24&71\\
\hline
Total&654&248&902\\
\hline
\end{tabular}
\caption{ Dataset of 902 patches from 19 repair tools from Wang et al. \cite{ASE20Wang}}	
\label{tab:dataset902}
\end{table}

\begin{table*}
\footnotesize
\renewcommand{\arraystretch}{2.5}
\centering
\begin{tabularx}{0.95\textwidth}{@{}Xrrrrrrrr@{}}
\hline
benchmark & \# original & \# duplicated & \# non-plausible & \# plausible & \# likely-correct  & \textbf{\# overfitting} & \textbf{ \makecell{\# correct \\(human-written)}} & \textbf{\#total} \\
\hline
Defects4J &37224& 19019 & 8491 & 9714  &4221 & \textbf{5493} &\textbf{ 798 } & \textbf{6291}\\

Bugs.jar &12419  & 8570 & 362 & 3487  & 1212 & \textbf{ 2275} & \textbf{986}&\textbf{\numprint{3261}} \\
Bears &13181 &10188 & 337 &2656  &2125 & \textbf{531} & \textbf{219}& \textbf{750}\\
\hline
Total & \numprint{62824} & \numprint{37777} & \numprint{9190} & \numprint{15857} & \numprint{7558} & \textbf{\numprint{8299}} & \textbf{\numprint{2003}} & \textbf{\numprint{10302}}\\
\hline
\end{tabularx}
\caption{Dataset of 10362 labelled patches from RepairThemAll \cite{repairthemall} per our automated labelling technique. This dataset is significantly larger than that of \autoref{tab:dataset902}.} 	
\label{tab:RepairThemAll}
\end{table*}


ODS uses supervised learning, which requires a corpus of labeled data used for training the model.
In this work, the training and testing data consists of patches, each patch with one particular label: \emph{correct} or \emph{overfitting}. 
We recall that overfitting patches are those patches generated by automatic program repair systems that pass the test suites provided by developers but do not actually fix the bug.

\TSErevision{
In this work, we consider two datasets of  patches:
the existing dataset from Wang et al.~\cite{ASE20Wang} with labeled patches, and the largest ever dataset from the RepairThemAll experiment \cite{repairthemall} with unlabeled patches.
}

\subsubsection{Wang et al.'s Dataset}
\label{sec:datasetwang}

\TSErevision{
Wang et al. \cite{ASE20Wang} provide a curated  dataset with 902 labeled patches for Defects4J bugs generated by 19 repair tools. 
\autoref{tab:dataset902} presents the details of this dataset, it contains 654 patches manually labeled as overfitting and 248 patches manually labeled as correct by the authors of the corresponding tools. We use this existing dataset to carry out a fair comparison between our work and the state-of-the-art, which has been used to evaluate \cite{Xiong-ACS-ICSE17,ASE20Wang}.
}

\subsubsection{RepairThemAll Dataset}
\label{sec:datasetrepairthemall}

\TSErevision{The existing datasets of manually labeled patches can be considered too small for machine learning:
Xiong et al.'s dataset has 139 labeled patches \cite{patchsim} from 6 repair tools,  Ye et al.'s dataset \cite{drr} has 628 labeled patches from 14 repair tools, and Wang et al.'s dataset \cite{ASE20Wang} contains 902 labeled patches, which is to know the largest to date in our knowledge. This is because the scale of the labeled patch dataset was limited by manual assessment. 
}

To overcome the absence of a large-scale labeled patches and the limitations of manual patch assessment at scale, 
we follow the methodology of Ye et al. \cite{drr} to assess patch correctness with \textbf{r}andomly generated tests based on \textbf{g}round \textbf{t}ruth patches (RGT).
This approach leverages RGT tests to differentiate program behavior between ground truth patches and program repair patches. 
The rule to classify an overfitting  patch is: \textit{if a patched program makes any RGT test fail, it is considered overfitting, otherwise likely-correct.}

We first gather patches from the recent experiment \textit{RepairThemAll} by Durieux et al. \cite{repairthemall}, which executed 11 repair systems on five bug benchmarks.
In particular, we focus exclusively on the three benchmarks which contain real-world bugs from large and open-source projects: Defects4J~\cite{defects4j},  Bugs.jar~\cite{bugjar} and Bears~\cite{Bears2019}.
However, RepairThemAll does not provide labels for the generated patches.
Therefore, to obtain the labels of patches, we reuse the existing RGT tests for Defects4J \cite{defects4j} generated by Evosuite \cite{evosuite},  and we generate additional Evosuite tests with 10 random seeds for Bugs.jar and Bears bugs.

\autoref{tab:RepairThemAll} presents the statistics of patches generated by RepairThemAll for three benchmarks. 
By analyzing the \numprint{62824} program repair patches (given in the second column), we discard
 \begin{inparaenum}[\it 1)]
\item \numprint{37777} duplicated patches generated by different repair systems (third column);
\item \numprint{9190} non-plausible patches that fail to pass the developer provided test cases (fourth column).
 \end{inparaenum}
We obtain \numprint{15857} plausible patches, shown in the fifth column, each of which succeeds in applying to the buggy program and passes all developer provided test cases.
The plausible patches are further labeled as likely-correct and overfitting according to the RGT test execution results. 
Eventually, we obtain \numprint{7558} labeled likely-correct patches and  \numprint{8299} labeled overfitting patches.

However, we do not consider the labeled likely-correct patches as the correct samples to train the learning model of ODS because their correctness is not guaranteed due to the incompleteness test cases ~\cite{drr,le:reliability-patch-assess}. 
That is, for example, a generated test may not include an input that exposes the overfitting patch.
Instead, we employ the human-written patches from three benchmarks as the correct samples, which are given in the eighth column called correct (human-written) patches.

The last three columns of \autoref{tab:RepairThemAll} describe the training set of ODS, including the overfitting patches, correct patches and total patches. 
Our dataset contains \numprint{8299} overfitting samples from 11 repair systems generated by \cite{repairthemall}: Arja~\cite{Yuan2017ARJAAR}, Cardumen~\cite{cardumen}, Dynamoth~\cite{dynamoth}, GenProg by Arja~\cite{Yuan2017ARJAAR}, jGenProg~\cite{martias2016defects4j}, jKali~\cite{martias2016defects4j}, jMutRepair~\cite{mutation},
Kali by Arja~\cite{Yuan2017ARJAAR}, Nopol~\cite{nopol}, NPEFix~\cite{durieuxNpeFix}, RSRepair by Arja~\cite{Yuan2017ARJAAR}.
Also, this dataset includes \numprint{2003} correct patches: 798, 986 and 219 human-written patches from Defects4J, Bugs.jar and Bears.
In total, we have collected \numprint{10302} labeled patches.

\subsection{Dataset Cleaning}

As mentioned in Section \ref{sec:featurs}, we extract features from the AST diff computed between the buggy program and the patched program.
We observe that those AST diff from patches have some noise, i.e., changes that are not related to repair the bug.

They could be: 
\begin{inparaenum}[\it 1)]
\item redundant code, e.g., Arja and jGenProg generate patches by adding redundant existing code;  
\item code optimizations added together with a bug fix in the human-written patches, which has been reported in \cite{drr};  
\item multiple-file modifications for fixing a single bug. We observe that a program repair patch can modify up to eight files while the bug only exists in one file and the ground truth patch (i.e., human-written patch) actually modifies only that buggy file.
\end{inparaenum}

These unrelated changes can lead to meaningless code feature extraction and affect the performance of classifications \cite{PLDI20TypeError,tse-18-dbn}.  
To prune noisy data, we use the Tukey method \cite{turkeymethod} to detect outliers. 
We detect outliers from the features of the numerical value.
We consider outliers as rows that have at least a specific number of outlied numerical values. In our study, we consider outliers as rows that have at least 15 outlied numerical values.

\subsection{Imbalanced Data Handling}
Imbalanced data occurs in all the three bugs benchmarks we considered, as shown in \autoref{tab:RepairThemAll}.  
Overall, there are \numprint{2003} correct patches and \numprint{8299} overfitting patches. The correct patches are around 20\% of the overfitting patches. 
The imbalanced data can lead to poor prediction
performance~\cite{icse'15-defect-predict}.
To overcome this problem, we perform the synthetic minority over-sampling technique SMOTE \cite{smote}, which has been widely used on the imbalanced source code learning tasks (e.g., \cite {tse-18-dbn, icse'15-defect-predict}). Specifically, we consider the \textit{minority} strategy of SMOTE which only resamples the correct patches to the same amount of overfitting patches.

\subsection{Terminology and Evaluation Metrics}
\label{sec:metrics}
Now, we determine the core evaluation terminology and evaluation metrics.
\begin{itemize}
\item{True Positive (TP): The prediction for a patch is true positive if an overfitting patch is rightly classified as overfitting.}
\item{True Negative (TN): The prediction for a patch is true negative if a correct patch is rightly classified as correct.}
\item{False Positive (FP): The prediction for a patch is false positive if a correct patch is wrongly classified as overfitting.}
\item{False Negative (FN): The prediction for a patch is false negative if an overfitting patch is wrongly classified as correct.}
\end{itemize}

\label{sec:roc}
\emph{Evaluation Metric} 
we consider precision (\autoref{equ:precision}), recall (\autoref{equ:recall}), and accuracy (\autoref{equ:accuracy}) to measure the effectiveness of ODS. These metrics have been widely adopted to evaluate classification models \cite{tse-18-dbn,icse'20CPCAC}.

\begin{equation}
 Precision = \frac{TP}{TP+FP}
 \label{equ:precision}
\end{equation}

\begin{equation}
 Recall = \frac{TP}{TP+FN}
 \label{equ:recall}
\end{equation}

\begin{equation}
 Accuracy = \frac{TP+TN}{TP+FP+TN+FN}
 \label{equ:accuracy}
\end{equation}

Precision evaluates the proportion of detected overfitting patches that are truly overfitting. 
A higher precision is demanded by program repair research as we do not want to discard correct patches. In our work, the higher precision shows higher reliability of the ODS model in discarding overfitting patches. 
However, comparing the overfitting patch prediction model by using only precision may be incomplete.
For example, one can only classify patches with higher confidence values as overfitting to achieve a higher precision score, which could result in a low recall score. 

Recall evaluates the effectiveness of a model to classify overfitting patches.
Higher recall is also critical for developers who do not want to waste their efforts on analyzing a large number of overfitting patches. The higher recall rate of the classification model, will bring a higher quality of patches delivered to developers and researchers. Comparing only the recall is also incomplete, as a model can predict all patches as overfitting to achieve 100\% of recall score.

To overcome the above issues, we also consider the accuracy to measure the performance of the patch correctness prediction, which is the comprehensive evaluation of all TP, FP, TN, and FN.

\TSErevision{
Moreover, we introduce CPR (an acronym for correct patch ratio) to analyze the ratio of correct patches before and after applying the automatic patch assessment technique (e.g., ODS) for a given repair tool. 
Before applying ODS, we compute the $CPR_{orig}$ (\autoref{equ:cprorig}) as the original correct patch ratio, i.e., the number of correct patches (\#C) over the total number of generated patches (\#Patches).
After applying automatic patch assessment techniques, we compute the $CPR$ (\autoref{equ:cpr}) as the number of truly correct patches (TN) over all correct patches considered by the patch assessment model, i.e., the sum of TN and FN.
Increasing the CPR metric is beneficial, because it shows to what extent the automatic patch assessment technique mitigates the low precision problem in program repair, and it would mean saving developers' time by avoiding presenting them with overfitting patches.
}

\TSErevision{
\begin{equation}
 CPR_{orig} = \frac{\#C}{\#Patches}
 \label{equ:cprorig}
\end{equation}
}

\TSErevision{
\begin{equation}
 CPR = \frac{TN}{TN+FN}
 \label{equ:cpr}
\end{equation}
}

\subsection{Experimental Methodology}

\begin{table*}[th!]
\small
\centering
\renewcommand{\arraystretch}{2.5}
\begin{tabular}{p{0.15\linewidth}p{0.05\linewidth}p{0.05\linewidth}p{0.05\linewidth}p{0.05\linewidth}p{0.1\linewidth}p{0.1\linewidth}p{0.1\linewidth}p{0.1\linewidth}}
\hline

Tools  & TP & FP & TN & FN & Precision & Recall & Accuracy  & CPR \\
\hline
PatchSim  & 249 & 51 & 186 & 392 & 83.00\% & 38.85\% &  49.54\%  & 32.18\% \\
\hline
SimFeatures & 583 & 87 & 161 & 71  &  87.01\% &     89.14\% &  82.48\%  & 69.40\% \\
\hline
ProphetFeatures & 585 & 73 & 175 & 69 & 88.90\% & 89.45\% & 84.25\% & 71.72\%
\\
\hline
ODS  &  620 & 66 & 182 & 34 & \textbf{90.38\%} & \textbf{94.80\%} & \textbf{88.91\%}  & \textbf{ 84.26\%} \\

\hline

\end{tabular}
\caption{\TSErevision{Comparison of the effectiveness of ODS with the state-of-the-art.}}	
\label{tab:rq1-new}
\end{table*}

\subsubsection{Protocol of RQ1}
\TSErevision{
In RQ1, we perform an experiment to compare the effectiveness of ODS with the state-of-the-art patch assessment techniques from Xiong et al. \cite{patchsim}, Wang et al. \cite{ASE20Wang}, and Long and Rinard \cite{prophet}. 
Xiong et al. \cite{patchsim} proposed a dynamic patch assessment technique based on the similarity of test executions before and after applying a patch, we refer to their technique as PatchSim in our paper. 
Wang et al. \cite{ASE20Wang} proposed overfitting patch detection based on 8 static code features, we refer to their technique as SimFeatures in our paper since their features focus on the similarity between the buggy context and the patch code.
Long and Rinard \cite{prophet} proposed an overfitting patch ranking technique called Prophet which learns from correct human-written patches. We note that transforming a ranking technique into a classification technique requires an arbitrary threshold, and studying the impact of the threshold is out of the scope of our paper. Thus, for a fair and meaningful comparison, we consider the Prophet features which are referred to as ProphetFeatures in our paper, and we put them in the same prediction algorithm as the others.
Please note that Prophet was originally designed for the C language and our experiment is based on patches generated for Java, thus we have re-implemented a Java version of Prophet.}

\TSErevision{
Per the best practices, all approaches are  evaluated on the same dataset, that one presented and described in Section \ref{sec:datasetwang}, composed of 902 patches. 
Since the feature extraction technique of SimFeatures is not publicly available, we compare our experimental results with the numerical results reported in their paper \cite{ASE20Wang}, which evaluated that approach on the same dataset (i.e., we have not re-executed SimFeatures).
We both train and test these ODS and ProphetFeatures with the same setup, per the setup of SimFeatures:
1) We use 10-fold cross-validation;
2) We use the same classifier (random forest).
To handle the imbalance of training samples, we apply SMOTE in each fold cross-validation. 
With this protocol, we aim at only measuring the effect of the set of features, and not that of the underlying machine learning algorithm. 
Finally, we compute the metrics discussed in Section \ref{sec:metrics} for the four techniques: precision, recall, accuracy and CPR.
}

\TSErevision{
Next, we investigate how ODS affects each repair tool individually.
We break up the 902 patches by 19 repair tools. 
For each tool, we use the patches from the other 18 repair tools as training data and the ones from the considered repair tool as testing. 
This can be considered as cross-validation over tools.
We compute and report on $CPR_{orig}$  and $CPR_{ODS}$ to show how ODS improves the performance of each tool. Recall that the CPR \autoref{equ:cpr} indicates the truly correct patch ratio over all patches generated by a tool. A higher CPR shows a higher precision of the patches provided to the developers. 
This means saving developers time in analyzing the overfitting patches and building trust for automatic program repair.
}

\subsubsection{Protocol of RQ2}
\label{sec:protocolrqone}

\TSErevision{In this RQ, we qualitatively compare the results from ODS and PatchSim in order to understand: 1)The complementary expressed as  the number of overfitting patches that ODS and PatchSim can detect; 2)The differences in execution time between a static  approach (ODS) and a dynamic approach (PatchSim).
Since the qualitative analysis of these two tools is time-consuming,
we evaluate the complementary of ODS against PatchSim on the same 139 patches used by Xiong et al. \cite{patchsim}.
Those patches form the testing set.
}

In this experiment, we use the combination patches from three benchmarks  presented in Section \ref{sec:datasetrepairthemall} (i.e., Defects4J, Bugs.jar and Bears) as a training set to build the ODS model.
To handle the imbalance of training samples, we apply SMOTE in the minority class.
However, as the training set contains patches of the same projects from the testing set (i.e., Lang, Chart, Math and Time), we need to discard all patches of those projects from the training set to ensure all testing samples are unseen by the ODS model.  
For example, Defects4J and Bugs.jar both contain correct and overfitting patches generated for the \emph{Math} project, then we discard all patches of \emph{Math} from the training set.  The reason we do this is to ensure all testing samples are unseen by the ODS model, because the patches from the same project may contain similar context. 
For instance, the human-written patches of Closure-62 and Closure-63 from Defects4J are actually identical. 
Specifically in this experiment, for the four considered projects of Defects4J, we create corresponding four pairs of training and testing. For instance, we test all patches from project $P$, where all patches from the testing project $P$ are discarded in the training set.
We record the number of the training set and testing set, and the results of precision, recall, and accuracy.


Moreover, we compare the runtime performance of ODS and PatchSim.
Specifically, we measure and compare the execution time used by these two techniques.  We execute ODS and PatchSim in the same environment\footnote{Intel(R) Core(TM) i5-6260U with 4 CPUs on the Ubuntu 16.04 operating system.} to classify the 139 patches from Xiong et al.'s dataset. We record the total prediction time and average prediction time per patch. 
Specifically for ODS, we also collect the feature extraction time and the total training time.

\subsubsection{Protocol of RQ3}
\TSErevision{In this RQ, we evaluate ODS on more projects to improve the external validity of our evaluation. 
We consider patches from the RepairThemAll experiment (Section \ref{sec:datasetrepairthemall}), grouping them by projects.
We recall that this dataset contains:
\begin{inparaenum}[a) ]
\item Automatic program repair patches classified as \emph{overfitting};  
\item The human-written patches are considered as \emph{correct}.
\end{inparaenum}
}
Those patches collected are from 95 projects: 17 projects from Defects4J, 6 projects from Bugs.jar and 72 projects from Bears. 
Due to the space limitation, we only consider the projects with at least five patches. 
This results in 26 projects for evaluation, including 8 projects with both correct and overfitting patches and 18 with only correct patches, i.e., \TSErevision{no overfitting patch is able to be found in those projects.} 
We create training and testing sets per project as follows.
We test all patches from project $P$, where all patches from the testing project $P$ are discarded in the training set to ensure the testing patches are new and unseen as we do in RQ2.

\subsubsection{Protocol of RQ4}
\TSErevision{
In this RQ, we analyze the importance of all ODS features.
For this, we train ODS with the whole \numprint{10302} labeled patches presented in Section~\ref{sec:datasetrepairthemall} 
with the goal of:
1) Performing a feature selection study and 
2) Analyzing the most important ODS features.
In the feature selection experiment, we conduct 10-fold cross validation to record the accuracy of ODS with different subsets of features. 
Specifically, we evaluate ODS with three individual groups of features, per their nature: ODS with 150 code description features, ODS with 26 repair pattern features and ODS with  26 contextual features (see Section~\ref{sec:ods_features}). 
Next, we select the top-K most important features based on the state-of-the-art feature selection approach F-score \cite{featureselection,fscore}, which reveals the discriminative power of each feature independently from others. 
We study accuracy with $K \in {50,100,150}$.
Lastly, we analyze the top 10 important features of ODS and measure how each code feature is negatively or positively correlated with the accuracy of ODS:
the negatively correlated features are those with the lower value among overfitting patches;
the positively correlated features are those with the higher value among the overfitting patches.
}

\subsection{Experimental Parameter Settings}
\label{sec:parameters}
In this study, we apply the same parameters to all experiments.
Since our purpose is not to find the best training or test set, we do not spend too much effort on well tuning the parameters of ODS. 
We consider outliers as rows that have at least 15 outlied numerical values, which results in 833 (i.e., 8\%) training samples are discarded if we consider the whole \numprint{10302} training samples together. 
We use the default parameters of XGBoost (i.e., \textit{learning\_rate} sets to 0.3 and \textit{max\_depth} sets to 6),  turning the \textit{gamma} to 0.5. 
To reproduce our experiment, we control the random state seed as 42 to all configurations where randomness is required.


\section{Experimental Results}


\subsection{RQ1: Effectiveness}
\label{sec:odsvspatchsim}

\begin{table*}[th!]
\centering
\footnotesize
\renewcommand{\arraystretch}{2}
\begin{tabular}{p{0.1\linewidth}p{0.04\linewidth}p{0.04\linewidth}p{0.07\linewidth}p{0.04\linewidth}p{0.04\linewidth}p{0.04\linewidth}p{0.04\linewidth}p{0.06\linewidth}p{0.06\linewidth}p{0.06\linewidth}p{0.06\linewidth}p{0.06\linewidth}}
\hline

Tools  & \#O  & \#C & $CPR_{orig}$ & TP &  FP & TN &FN  & Precision & Recall & Accuracy  & $CPR_{ODS}$ \\

\hline
Arja & 52 & 5 & 8.77\% & 50 & 3 & 2 & 2 & 94.34\% & 96.15\% & 91.23\%  & 50.00\% \\
\hline
ACS & 7 & 15 & 68.18\% & 4 & 6 & 9 & 3 & 40.00\% & 57.14\% & 59.09\% & 75.00\%  \\
\hline
AVATAR & 38 & 19 & 33.33\% & 37 & 5 & 14 & 1 & 88.10\% & 97.37\% & 89.47\% & 93.33\% \\
\hline
CapGen & 41 & 25 & 37.88\%  & 34  & 4 & 21 & 7  & 89.47\% & 82.93\% & 83.33\%  & 75.00\% \\
\hline
Cardumen  & 10 & 2 & 16.67\% & 10 & 0 & 2 & 0 & 100.00\% & 100.00\% & 100.00\%  & 100.00\%  \\
\hline
DynaMoth & 21 & 1 & 4.55\% & 21 & 1 & 0 & 0 & 95.45\% & 100.00\% & 95.45\% & 0.00\% \\
\hline
FixMiner & 20 & 12 & 37.5\% & 18 & 1 & 11 & 2 & 94.74\% &  90.00\% & 90.63\% & 84.62\%   \\
\hline
GenProg & 45 & 5 & 10.00\% & 41 & 3 & 2 & 4 & 93.18\% & 91.11\% & 86.00\% & 33.33\%  \\
\hline
Jaid & 41 & 40 & 49.38\% & 35 & 19 & 21 & 6 & 64.81\% & 85.37\% & 69.14\% & 77.78\%  \\
\hline
jMutRepair & 17 & 5 & 22.73\% & 13 & 0 & 5 & 4 & 100.00\% & 76.47\% & 81.82\% & 55.56\% \\
\hline
Kali & 82 & 6 & 6.82\% & 70 & 4 &2 & 12 & 94.59\% & 85.36\% & 81.82\% & 14.29\%  \\
\hline
kPAR & 52 & 10 & 16.13\% & 48 & 1 & 9 & 4 & 97.96\% & 92.31\% & 91.94\% &69.23\% \\
\hline
Nopol & 30 & 1 & 3.22\% & 17 & 0 & 1 & 13 &  100.00\% & 56.67\% & 58.06\% & 7.14\% \\
\hline
RSRepair  & 39 & 2 & 4.88\% & 38 & 1 & 1 & 1 & 97.44\% & 97.44\% & 95.12\% & 50.00\% \\
\hline
SequenceR & 56 & 17 & 23.29\% & 18 & 0 & 17 & 38 &  100.00\% & 32.14\%  & 47.95\% & 30.91\% \\
\hline
SimFix & 45 & 22 & 32.84\% & 18 & 3 & 19 & 27   & 85.71\% & 40.00\%  & 55.22\% & 41.30\%  \\
\hline
SketchFix & 9 & 16 & 64.00\%  & 6 & 3 & 14 & 3 & 75.00\% & 66.67\% & 80.00\% & 82.35\%  \\
\hline
SOFix & 2 & 21 & 91.30\% & 2 & 4 & 17 & 0 & 33.33\% & 100.00\% & 82.61\%  &  100.00\%   \\
\hline
TBar & 47 & 24& 33.33\% & 40 &2 & 22 & 7 & 95.24\% & 85.11\% & 87.32\% & 75.86\%   \\
\hline

AVG & - & - & 29.73\% & -  &- &-  & - &  86.28\% & 80.64\% & 80.33\% & 58.72\% \\
\hline
\end{tabular}
\caption{\TSErevision{The effectiveness of ODS to detect overfitting patches and to improve the correct patch ratio per tool.}}	
\label{tab:rq1-by-tool}
\end{table*}

\TSErevision{
We compare the effectiveness of ODS with the state-of-the-art patch assessment technique PatchSim \cite{patchsim}, SimFeatures \cite{ASE20Wang} and ProphetFeatures \cite{prophet} on 902 patches presented in \autoref{tab:dataset902}.
\autoref{tab:rq1-new} gives the detailed results.
We present the prediction results in the second to the ninth columns, including the number of true positives, false positives, true negatives and false negatives, as well as the evaluation metrics of precision, recall, accuracy and correct patch ratio (CPR).
The best evaluation metrics in these four techniques are shown in bold.
Please note that PatchSim has a slightly fewer total number of patches, because PatchSim does not terminate for some patches due to lack of heap space when a large number of tests cover the patched method \cite{ghanbari2020validation,ASE20Wang}.
}

\TSErevision{
\autoref{tab:rq1-new} yields two main findings. 
First, ODS outperforms PatchSim, SimFeatures and ProphetFeatures in identifying overfitting patches in all evaluation metrics: the precision, recall, accuracy, and CPR are better.
ODS is able to identify 620 out of 654 overfitting patches, and  also correctly classifies 182 out of 284 correct patches. This leads to 90.38\%, 94.80\% and 88.91\% in precision, recall and accuracy, respectively.
Second, the biggest improvement made by ODS is on the correct patch ratio (CPR).
Per this evaluation metric, PatchSim, SimFeatures and ProphetFeatures has a respective performance of 32.18\%, 69.40\% and 71.72\%, and ODS bumps it to 84.26\%. This means that more correct patches are presented to developers, and that ODS is considered the best technique to mitigate the low precision patch generation problem in program repair.
}

\TSErevision{
\textbf{Comparison between ODS and PatchSim.}
Per accuracy, ODS outperforms PatchSim by 39.37\% (88.91\% versus 49.54\%). This is because ODS is able to identify 94.80\% of overfitting patches, i.e., has a high recall, while PatchSim only detects 38.85\% of overfitting patches. 
Per false positives, PatchSim generates fewer false positive cases than ODS: it has 51 false positive cases, where ODS yields 66 false positive cases, because it was carefully designed and optimized for minimizing the number of false positives. 
However, PatchSim fails to identify 392 overfitting patches (i.e., false negative cases). This results in a lower correct patch precision, where 32.18\% of patches are truly correct. 
On the contrary, ODS only yields 34 false negative cases,  which leads to  84.26\% in CPR. 
In summary, despite a higher false positive rate, ODS is able to identify more overfitting patches than PatchSim, and achieves a significantly higher CPR. This is important for practitioners who could have more confidence in the patches labeled as correct.}

\TSErevision{
\textbf{Comparison between ODS, SimFeatures and ProphetFeatures.}
Recall that ODS, SimFeatures and ProphetFeatures are three different types of static code features.
For accuracy, ODS outperforms SimFeatures (88.91\% versus 82.48\%) and ProphetFeatures  (88.91\% versus 84.25\%).
For false positives, ODS outperforms SimFeatures by having 21 fewer false positive cases (66 by ODS and 87 by SimFeatures), and ODS outperforms ProphetFeatures by having 7 fewer false positive cases less (66 by ODS and 73 by ProphetFeatures).
Moreover, ODS detects less than half false negative cases than the other two techniques: ODS produces 37 fewer cases than SimFeatures (34 versus 71) and 35 fewer cases than ProphetFeatures (34 versus 69).  
This allows ODS to achieve a higher CPR: 84.26\% for ODS while SimFeatures and ProphetFeatures achieve 69.40\% and 71.72\%, respectively.
This experiment clearly shows that ODS' feature set improves over the state-of-the-art features of SimFeatures \cite{ASE20Wang} and ProphetFeatures \cite{prophet}.
This is explained by both the number and the types of features. 
First, ODS considers more features: 202 features versus 8 features in SimFeatures, 35 atomic features and 5 modification features in ProphetFeatures (the original Prophet features before crossing them together). 
Second, ODS has more diverse features, it considers syntactic features (code description features), semantic features (repair pattern features), and contextual features. 
On the contrary, in SimFeatures, Wang et al. \cite{ASE20Wang} mostly focus on code similarity features, which is a subset of ODS's  contextual features. 
In ProphetFeatures, Long and Rinard \cite{prophet} consider atomic code features and modification features, which also have been considered and extended in ODS's code description features. For example, ProphetFeatures does not consider code removal in their modification features, while ODS detailed code removal features, e.g., \textit{removePartialIf} and \textit{removeWholeBlock}.
}

\TSErevision{
\textbf{Effectiveness of ODS per repair tool.}
\autoref{tab:rq1-by-tool} shows the effectiveness of ODS to classify overfitting patches for the 19 individual repair tools on 902 patches. The row are sorted by alphabetic order of name of the repair tools.
The last row gives the average evaluation metrics from 19 repair tools.
For example, per \autoref{tab:rq1-by-tool},  CapGen generates 41 overfitting patches and 25 correct patches for Defects4J. On those patches, ODS rightly detects 34/41 overfitting patches, 21/25 correct patches, and has only 4 false positives and 7 false negatives. This results in  89.47\%, 82.93\%, 83.33\% and 75.00\% in precision, recall, accuracy and CPR.}

\TSErevision{
Over the 19 repair tools, ODS achieves an average of 86.28\%, 80.64\% and 80.33\% in precision, recall and accuracy, respectively (bottom line).
Per precision,
ODS achieves 100.00\% precision for 4/19 repair tools, i.e., it does not misclassify a single correct patch, which is arguably helpful for developers.
ODS achieves more than 90\% precision in 8/19 repair tools, i.e., the false positive rate is less than 10\%.
ODS achieves more than 80\% precision in 3/19 repair tools, while
ODS achieves less than 80\% precision in 4/19 repair tools.
Per recall, ODS achieves more than 50\% recall in  17/19 repair tools, this shows ODS is able to help the majority of repair tools to filter out more than half of the overfitting patches.
}

\TSErevision{
\textbf{Correct patch ratio.}
ODS enables repair tools to increase the precision of generated patches, i.e., to present more correct patches to developers.
Individually, we observe that ODS positively increases the CPR for 18 of 19 repair tools, ranging from 7.14\% to 100.00\%.
For example, without ODS, \textit{Cardumen} original produces a correct patch ratio of 16.67\%, yet the correct patch ratio grows to 100.00\% after ODS is employed as post-filter.
There is one case that ODS decreases the original correct patch ratio for the repair tool \textit{DynaMoth}.
This is because \textit{DynaMoth} only produces one correct patch and ODS misclassifies it as overfitting, thus the CPR with ODS for DynaMoth is 0.00\%, which can be considered as an artefact of the measurement.
Overall, ODS helps 18/19 repair tools to increase the ratio of correct patches presented to developers: this would potentially improve the practical usage of program repair: save developers' time in analyzing overfitting patches and increase trust in automated program repair.
}

\begin{mdframed}
\TSErevision{
Answer  to  RQ1:  On the testing set of 902 patches generated by 19 repair tools,  ODS outperforms the state-of-the-art patch assessment technique PatchSim~\cite{patchsim}, SimFeatures~\cite{ASE20Wang}, and ProphetFeatures \cite{prophet}.  ODS produces fewer false positives, and is able to increase the correct patch ratio for 18/19 repair tools.
}
\end{mdframed}


\subsection{RQ2: Complementarity with PatchSim}
\TSErevision{
Now, we make a detailed analysis of static patch assessment (ODS) versus dynamic patch assessment (PatchSim).
\autoref{tab:rq1} presents the number of classified patches by ODS and PatchSim.
The first three columns give the information of the training set and the fourth to sixth columns present the testing set statistics.
In the first column, \emph{All - P} indicates all training patches from \autoref{tab:RepairThemAll} minus the patches from P, where P is the project in the testing set. 
The number of overfitting and correct patches are respectively denoted as \textit{\#O} and \textit{\#C}.
Those are the original training number of patches before applying SMOTE.
Then, we present the PatchSim prediction results in the seventh to the tenth columns.
We give the same prediction results by ODS in the last four columns.  
}
\TSErevision{
For example, the second row indicates that, by using \numprint{7267} overfitting patches and \numprint{1940} correct patches from all collected patches minus the \textit{Lang} patches as the training set, ODS rightly detects 10 of 11 overfitting patches, 4 of 4 correct patches, with 0 false positive and 1 false negative (it fails to identify 1 overfitting patch). 
Now we analyze the complementary between ODS and PatchSim, as well as the runtime performance of these two techniques.}

\begin{table}[th!]
\scriptsize
\centering
\renewcommand{\arraystretch}{2.5}
\begin{tabular}{p{0.13\linewidth}p{0.025\linewidth}p{0.04\linewidth}|p{0.045\linewidth}p{0.02\linewidth}p{0.022\linewidth}|p{0.01\linewidth}p{0.007\linewidth}p{0.007\linewidth}p{0.022\linewidth}|p{0.007\linewidth}p{0.007\linewidth}p{0.007\linewidth}p{0.007\linewidth}}
\hline
\multicolumn{3}{c}{Training}& \multicolumn{3}{c}{Testing} &  \multicolumn{4}{c}{PatchSim } & \multicolumn{4}{c}{ODS } \\
\hline
Proj. & \#O& \#C &Proj. & \#O & \#C   
&TP& FP&TN& FN  & TP& FP&TN& FN \\
\hline

All - Chart & 6977& 1978& Chart &23 & 3 
& 14& 0& 3& 9  
&14 &0 &3 &9  \\

All - Lang &7267& 1940& Lang &11&4 
&6& 0& 4& 5  
&10& 0& 4 & 1  
\\

All - Math & 3625&1783 & Math & 63& 20 & 
33& 0 & 20 & 30  
&35 &4 &16 &28  \\

All - Time &8299 & 1974& Time &13 & 2  
&9 &0 & 2& 4 
& 11& 1& 1 & 2  
\\
\hline


\end{tabular}
\caption{Comparison between ODS and PatchSim on Xiong et al.'s dataset of 139 patches  (All - P = all training patches minus patches from project P, O=Overfitting patches, C=Correct patches).
}	
\label{tab:rq1}
\end{table}

\begin{figure}[th!]
  \centering
\includegraphics[width=0.5\textwidth]{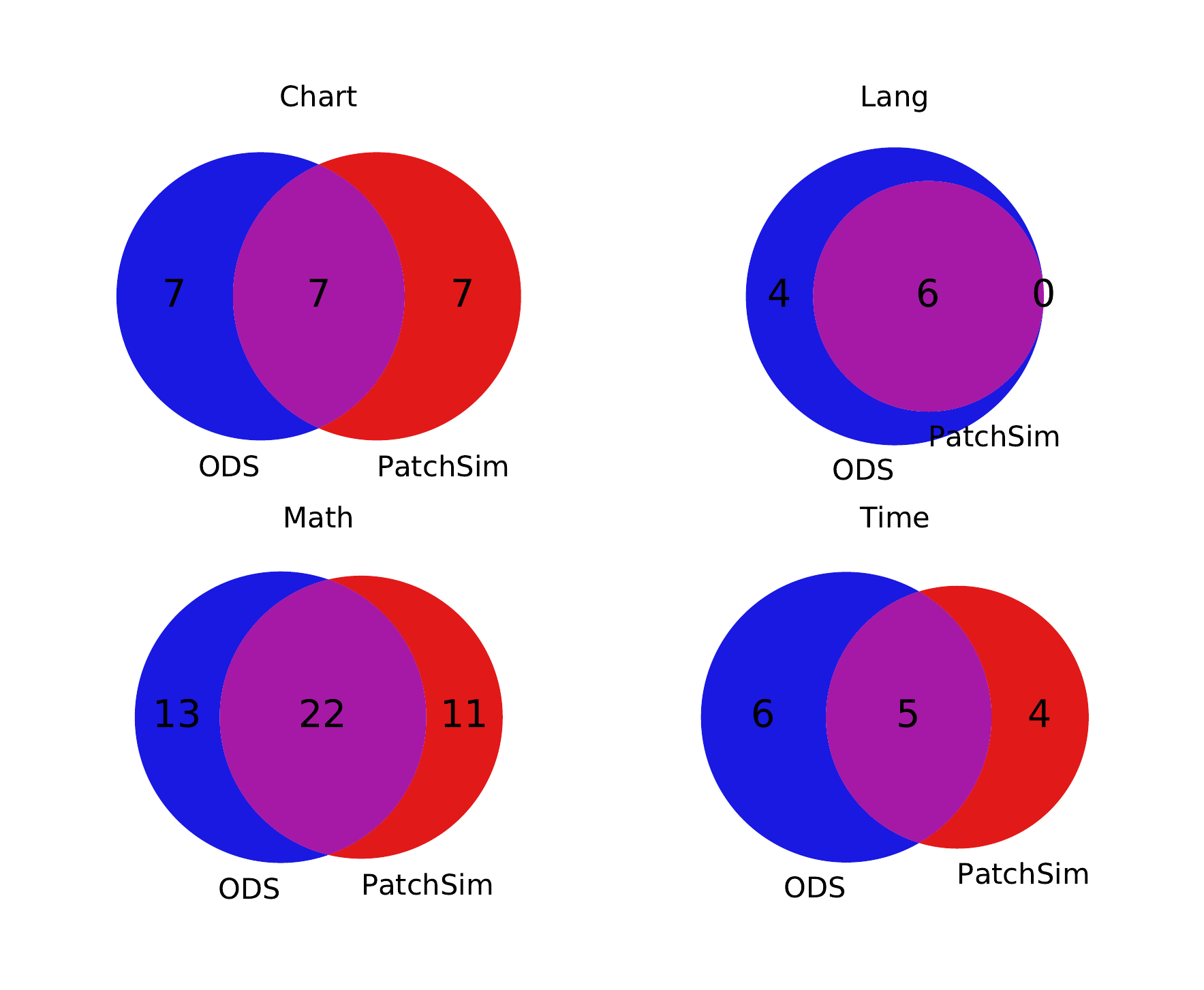}
\caption{Intersection of the detected overfitting patches (true positives) by  ODS and PatchSim.}
\label{fig-venn-ods-patchsim}
\end{figure}

\textbf{Complementary between ODS and PatchSim.}
\autoref{fig-venn-ods-patchsim} shows the intersection of correctly detected overfitting patches (i.e., the true positive cases) between ODS and PatchSim in four projects. 
The blue and red circles represent ODS and PatchSim, respectively.
For example, for \textit{Chart} project (top left diagram), ODS and PatchSim detect 7 overfitting patches in common and they individually detect 7 different overfitting patches. 
\TSErevision{Considering four projects together, our experiment shows that PatchSim and ODS together identified 92 patches, and 40 of them can be identified by both techniques. The remaining 52 patches can only be identified by either ODS or PatchSim.}

Notably, ODS and PatchSim do not discard the same overfitting patches. They complement each other in three projects (\textit{Chart}, \textit{Math} and \textit{Time}) and overlap in one project (\textit{Lang}). 
This is because these two techniques are based on different strategies. PatchSim is a dynamic patch assessment technique that analyzes execution traces between passing tests and failing tests. 
ODS, on the contrary, is a static technique that builds a probability model to predict the patch's correctness by considering code features.  
Since they use strategies of different natures, they have different results.

\textbf{Case study of a true positive case.}
In \autoref{lst:chart17}, we give an example of an overfitting patch generated for bug Chart-17, which is detected as overfitting by ODS but not PatchSim.  
Listing 1a shows the overfitting patch by Nopol for this bug\footnote{which is identified as Patch90 in our dataset}, and Listing 1b gives the human-written patch.
The program repair patch and the human-written patch are different to a large degree yet PatchSim considers it as correct based on the test execution comparison: the passing tests perform similarly and the failing tests perform differently after patching the bug. 
ODS well considers it as overfitting because of code features related to two global variables (\textit{this.data} and \textit{this.range}) that are used in conditional expressions.  
For the ODS learned model, the local variables are more likely to be used in a correct patch, rather than global variables, particularly when local variables are around the patched code (e.g., variables $end$ and $start$).
Consequently, ODS identifies it as an overfitting patch. 
\begin{listing}[th!]
\noindent\begin{minipage}[b]{0.48\textwidth}
    \subcaption{An overfitting patch for Chart-17 from Patch90 by Nopol}
    \begin{lstlisting} [firstnumber=419] 
 if (end < start) {
+  if (this.data.size() == this.range.length()){
    throw new IllegalArgumentException("..");}
+  }
    \end{lstlisting}
    \label{chart17-program-patch}            
    \end{minipage}%
    \hfill
    \begin{minipage}[b]{0.48\textwidth}
    \subcaption{A human-written patch for Chart-17 }
    \begin{lstlisting}[firstnumber=419] 
public Object clone() {
- Object clone = createCopy(0, getItemCount()-1);
+ TimeSeries clone = (TimeSeries) super.clone();
+ clone.data = ObjectUtil.deepClone(this.data);
    \end{lstlisting}

    \label{chart17-human-patch}            
\end{minipage}%
\caption{An overfitting patch only detected by ODS and the corresponding human-written patch.}
\label{lst:chart17}
\end{listing}


\begin{listing}[t!]

\noindent\begin{minipage}[b]{0.45\textwidth}
    \subcaption{Patch by HDRepair and  jGenProg for Math-70}
    \begin{lstlisting} [firstnumber=419] 
-  return solve(min, max);
+  return solve(f, min, max);
    \end{lstlisting}
    \label{patch-Lang-43-ground-truth}            
    \end{minipage}%
    \hfill
    \begin{minipage}[b]{0.45\textwidth}
    \subcaption{Patch by jGenProg for Math-73}
    \begin{lstlisting}[firstnumber=419] 
-  return solve(f, min, yMin, max, yMax, initial, yInitial);
+  return solve(f, min, max);
    \end{lstlisting}

    \label{patch-Lang-43-1}            
\end{minipage}%
\caption{False positive cases of Math-70 and Math-73}
\label{lst:fpmath70and73}
\end{listing}

\textbf{Case study of a false positive case.}  We manually investigate the causes of those false positive cases.
\autoref{lst:fpmath70and73} presents three patches\footnote{Two patches generated for Math-70 by HDRepair and jGenProg are identical} generated for two bugs Math-70 and Math-73 bugs that are identical to the human-written patches, yet wrongly classified as overfitting by ODS.  
We note that the three patches use the same repair pattern: they fix the bug by replacing a method call by its overridden version. However, ODS does not have a feature to capture the overridden method replacement.   
This shows that ODS is interpretable and traceable: one can understand the reasons behind a wrong classification and add the missing features that could potentially mitigate the problem.

\textbf{Runtime Performance.}
\autoref{tab:runtime-performance} summarizes our observations about the runtime performance.
The first column presents the two patch assessment techniques, and the second column gives the total time spent on the feature extraction and training process. Since PatchSim does not require such a process, we put a '-' in the corresponding cell.
The third column presents the total prediction time spent in classifying the correctness of 139 patches of RQ1. 
The fourth column gives the average prediction time per patch.

ODS spent approximately 10 hours for feature extraction and training. Most of the time is spent on the feature extract phase and the training time takes less than 10 minutes.  Please note that this feature extraction phase only needs to be done once for a given training dataset and the learned model can be reused. 
PatchSim takes approximately 70 hours to classify all 139 patches, while ODS takes less than 30 minutes for predicting the correctness of all patches.  
On a per-patch basis, PatchSim takes on average 30 minutes and ODS takes on average 13 seconds for classifying a single patch. 
Therefore, ODS is 138 times faster than PatchSim in identifying overfitting patches. Yet, this comes to the cost of 10 hours to extract code features and train the model.  The total time of the experiment is 10.5 hours for ODS (10 hours for features extraction and training + 30 minutes for prediction), which is also significantly faster than PatchSim (70 hours).

\begin{mdframed}
\TSErevision{
Answer to  RQ2: 
Our in-depth analysis shows that ODS and dynamic patch assessment technique PatchSim do not detect the same overfitting patches, and thus they can be considered complementary.
On the runtime side, ODS is 138 times faster than PatchSim for classifying a single patch as overfitting or not (13 seconds versus 30 minutes). 
The main reason is that ODS is a purely static approach, which saves compilation, instrumentation, and execution time. 
For the practitioners who use program repair tools, it is significant: they get an overfitting diagnostic almost as soon a patch is generated.
}
\end{mdframed}

\begin{table}[t!]
 \centering
 \begin{tabular}{@{}lccc@{}}
  \hline
\multirow{2}{*}{Approach} & Feature & Total  & Prediction  \\
 & + Training &  Prediction & Time/Patch     \\
\hline
PatchSim & - & $\approx$ 70 hours & $\approx$ 30 mins \\
ODS  & $\approx$ 10 hours  & $<$30 mins  & 13 secs  \\
 \hline
 \end{tabular}
 \caption{Comparison of runtime performance of PatchSim and ODS.}	
\label{tab:runtime-performance}
\end{table}

\subsection{RQ3: Project Independence}
\begin{table*}[t!]
\footnotesize
\renewcommand{\arraystretch}{1.9}
\centering
\begin{tabular}{p{0.08\linewidth}|p{0.04\linewidth}p{0.04\linewidth}p{0.05\linewidth}|p{0.04\linewidth}p{0.04\linewidth}p{0.05\linewidth}|p{0.04\linewidth}p{0.04\linewidth}p{0.04\linewidth}p{0.04\linewidth}p{0.06\linewidth}p{0.06\linewidth}p{0.06\linewidth}}
\hline

 \multirow{2}{*}{Project P}&\multicolumn{3}{c|}{Training (discarded from P)} & \multicolumn{3}{c|}{Testing on P} &\multicolumn{7}{c}{Results}\\
  & \# O& \# C & \# Sum  & \# O & \# C & \# Sum  & TP & FP& TN &FN & Precision & Recall & Accuracy \\
\hline
Math& 3625 & 1783  & 5408 & 4674 & 220 & 4894 & 3056  & 48 & 172 & 1618 & 98.5\% & 65.4\% & 66.0\%\\
Chart & 6977 & 1978  &8955   & 1322 & 25 & 1347 & 1026  & 7 & 18 & 296    & 99.3\% & 77.6\%  & 77.5\%\\
Lang & 7267 & 1940 & 9207 & 1032 & 63  & 1095   &767  & 24 & 39 & 265    & 97.0\% & 74.3\%  & 73.6\%\\
Jackrabbit & 7950 & 1761 & 9711 & 349 & 242 & 591 & 248  & 44 & 198 & 101 & 84.9\%  & 71.1\% & 75.5\% \\
Flink  & 7949 & 1937 & 9886 & 350 & 66 & 416 &244  & 13 & 53 & 106 & 94.9\% & 69.7\%  & 71.4\% \\
Accumulo & 8022 & 1922 & 9944  & 277 & 81 & 358 & 225  & 12 & 69 & 52 & 94.9\% & 81.2\%  & 82.1\% \\
Traccar & 8132 & 1962 & 10094 & 167 & 41 & 208 & 120  & 2 & 39 & 47 & 98.4\% & 71.9\%  & 76.4\%\\
Libra & 8171 & 2002  & 10173 & 128 & 1 & 129  & 122 & 0 &  1 & 6 & 100\% & 95.3\% & 95.3\%\\

\hline

 Wicket & 8299 & 1758 & 10057 & 0 & 245 & 245 & - & 35 & 210 & - & - & - & 85.7\% \\
 Closure & 8299 & 1831 & 10130 & 0 & 172 & 172 &-  & 57 & 115 & - & - & - & 66.9\% \\
 Camel & 8299 & 1879 & 10178 & 0 & 124 & 124 &- & 19 &  105 & - &  - & - & 84.7\% \\
 Jsoup & 8299 & 1915 & 10214 & 0 & 88 & 88 & - & 20 & 68 & - & - & - & 77.3\% \\
Log4J & 8299 &  1931 & 10230 & 0 & 72 &72 & - & 6 & 66 & -  & - & - & 91.7\% \\
Spoon & 8299 & 1953 &  10252    & 0 & 50 & 50  & - & 1 & 49 & - & - & -  & 98.0\% \\
Maven & 8299 & 1961 & 10260 & 0 &42  & 42 & - & 3 & 39 & - & - & - & 92.9\% \\
Mockito & 8299 & 1968 & 10267  & 0 & 35 & 35 & -  & 8 & 27 & -& - & - & 77.1\% \\
Time & 8299 & 1974& 10273  & 0 & 29 & 29 & - & 8 & 21  & - & - & - & 72.4\% \\
Cli & 8299 & 1975& 10274  & 0 & 28 & 28 & - & 8 & 20  & - & - & - & 71.4\% \\
JacksonXml & 8299 & 1985 & 10284 & 0 & 18 & 18 &- & 2 & 16 & -&- & - & 88.9\% \\
Spring & 8299 &  1986 & 10285 & 0 & 17 & 17 & - & 0 & 17 & - & - & - & 100\% \\
Codec & 8299 &  1986 & 10285 & 0 & 17 & 17 & -& 2 & 15 & -& -& - & 88.2\% \\
Csv & 8299 &  1986 & 10285 & 0 & 17 & 17 & - & 3 & 14 & -& - & - & 82.4\% \\
Gson &  8299 & 1988 & 10287 & 0 & 15 & 15 & - & 3 & 12 & -& - & - & 80.0\% \\
Incubator & 8299 & 1996 & 10295 & 0 & 7 & 7 & - & 1 & 6 & - & - & - & 85.7\% \\
Fresco & 8299 & 1998 & 10297 & 0 & 5 & 5 & - & 0 & 5 & - & - & - & 100\% \\
Molgenis& 8299 & 1998 & 10297 & 0 & 5 & 5 & - & 1 & 4 & - & - & - & 80.0\% \\
\hline
 \multicolumn{4}{c}{Total 26 projects}   & 8299 &  1725  &  10024 &  5808  &  327 & 1398 & 2491 & 94.7\%  & 70.0\% & 71.9\% \\
\hline
\end{tabular}
\caption{ Accuracy of ODS in a cross-project setting  (O=Overfitting patches, C=Correct patches). }	
\label{tab:rq3}
\end{table*}


We look at whether the overfitting knowledge gained from one project is useful for classifying patches generated from other projects.
\TSErevision{\autoref{tab:rq3} presents the ODS effectiveness in classifying \numprint{10024} considered patches from 26 projects that have more than five patches in our dataset.}
The projects are sorted by the total number of testing patches in descending order.
The first column gives the name of the project.
The second to fourth columns present the number of overfitting patches, correct patches, and the total number of samples in the training set. 
\TSErevision{Please note that those are the original training patch numbers before running SMOTE.}
The fifth and seventh columns show the testing patches from the evaluated projects. 
We present the results of TP, FP, TN, FN, precision, recall, and accuracy in the last seven columns. 
\TSErevision{The first eight projects contain both overfitting and correct patches while the latter 18 projects contain only correct patches. Note that per the literature, we know that some projects in the latter category have overfitting patches (e.g., Closure), yet per the labeling technique used in our paper, no test can be generated to show that they are actually overfitting, hence $\#O = 0$ for column in testing.}

This results in  TP, FN, precision and recall cannot be computed for a single class sample (i.e., projects only with correct patches), thus we put a '-' in the corresponding cells. 
We present the results by considering a total of all 26 projects in the last row.

For example, the first row indicates when ODS is evaluated for testing all patches for the \textit{Math} project, trained on \numprint{3625} overfitting patches and \numprint{1783} correct patches from all projects except those from \textit{Math} (note that this training set is the same as the one for testing \textit{Math} patches in RQ1, see the third row of \autoref{tab:rq1}) and tested on \numprint{4674} overfitting patches and 220 correct patches from \textit{Math}.  
ODS rightly detects \numprint{3056}/\numprint{4674} overfitting patches, 172/220 correct patches, it fails to identify \numprint{1618} overfitting patches and with 48 false positives. 
This means that for the \textit{Math} project, ODS achieves 98.5\%, 65.4\% and 66.0\% in precision, recall and accuracy, respectively.

Considering a total of \numprint{10024} patches from 26 projects, ODS rightly detects \numprint{5808} of \numprint{8299} overfitting patches, \numprint{1398} of \numprint{1725} correct patches, it fails to identify \numprint{2491} overfitting patches and with 327 false positive cases.  Overall, ODS achieves the scores of 94.7\%, 70.0\% and 71.9\% in precision, recall and accuracy, respectively. 

Those results are consistent with those presented in RQ1, where ODS produces respective precision, recall and accuracy of 93.3\%, 63.6\% and 67.6\% in the evaluation of 139 patches from 4 projects.
Since this experiment is made on a dataset two orders of magnitude bigger (\numprint{10024} versus 139), the external validity of the performance evaluation of ODS is significantly improved.

We note that ODS is generally effective on all projects and its effectiveness is not tied on specific projects. 
Considering overfitting patches only, ODS classifies more true positive cases than false negative cases in all 26 projects.
In total, \numprint{5808} overfitting patches are rightly classified and \numprint{2491} overfitting patches are wrongly classified.  
Considering correct patches only, ODS classifies more true negative cases than false positive cases in all 26 projects. 
In total, \numprint{1398} true negatives and 327 false positives are produced.  

Considering all patches together, ODS achieves an accuracy higher than 70\% in 24/26 projects.
Recall that we make sure that the training dataset does not contain any patch from the project considered for testing, this shows that the knowledge captured on one project can be applied to a new and unseen  project. 
\TSErevision{To our knowledge, this transfer effect has only been studied once by Long and Rinard \cite{prophet}. Since their paper focuses on patch ranking (and not patch classification), we are the first to measure and report on project independence for overfitting patch classification, and to show that the overfitting knowledge gained on one project can be applied to another project.
}
Also, there are 22 (except for the 4 projects evaluated in RQ1) new projects that have never been evaluated in the literature.
This transfer effect is important for ODS to be generally applicable in practice, because when practitioners start new projects, they do not have a history of program repair patches. 
To sum up, practitioners can directly reuse ODS to do patch classification when they deploy automatic program repair.

The performance variation of overfitting detection over projects is small but noticeable. There are two projects that achieve accuracy of 100\%, four projects achieve accuracy above 90\%, nine projects achieve accuracy between 80\% and 90\%, nine projects achieve accuracy between 70\% and 80\%, and two projects achieve accuracy between 60\% and 70\%.
This is because different projects have different characteristics in terms of coding style and complexity of the human-written patch considered in the testing dataset.
For example, we note the lowest accuracy of projects with only correct patches is 66.9\% for project \textit{Closure} which is still arguably high.  
Our  manual analysis shows that the human-written patches for \textit{Closure}  are complex because they consist of more than 10 lines of changed code, which is typically more than the training correct samples.

\begin{mdframed}
Answer  to  RQ3:  
Over a novel dataset of \numprint{10024} patches specifically designed for overfitting research, ODS reaches an overall accuracy  of 71.9\% (min accuracy 66.0\%, median accuracy 81.1\% and maximum accuracy 100\% from 26 projects). Per our original methodology ensuring that training patches come from different projects, this clearly shows that ODS can be used for overfitting detection on new projects.
Also this experiment made on a novel dataset: 1) confirms the performance of ODS per the experiments of RQ1; 2) improves its external validity since the dataset is an order of magnitude bigger.
This result is important for practitioners, it means they can directly use ODS to do patch classification on their projects with good performance, without any new data collection or new patch labeling for this particular project.
\end{mdframed}

\subsection{RQ4: Feature Analysis}
\label{sec:odsfeaturesinterpretability}

\begin{table}[th!]
\footnotesize
\centering
\renewcommand{\arraystretch}{2.5}
\begin{tabular}{p{0.32\linewidth}p{0.06\linewidth}p{0.06\linewidth}p{0.06\linewidth}p{0.06\linewidth}p{0.1\linewidth}}
\hline

Features  & TP & FP & TN & FN &  Accuracy  \\
\hline


ODS@CodeDescription & 7296 & 466  & 1537 &  1003 &  85.74\%   \\
ODS@RepairPattern & 5535 & 595 &1408 &2764 &  67.34\% \\
ODS@Contextual & 7681 & 480 &  1523 & 618 &  89.34\%  \\
\hline
ODS@50Best  & 7185 & 566 & 1437 & 1114 & 83.69\% \\
ODS@100Best & 7972 & 567 & 1436 & 327 & 91.32\% \\
ODS@150Best & 8047 & 527 & 1476 & 252 & 92.44\% \\
\hline
ODS@All & 8098 & 452 & 1551 & 201 & 93.66\% \\

\hline

\end{tabular}
\caption{\TSErevision{The impact of subsets of ODS features on performance.}}	
\label{tab:feature-selection}
\end{table}

\TSErevision{
\autoref{tab:feature-selection} presents the results of our feature selection study with different set of features to classify \numprint{10302} labeled patches of dataset RepairThemAll (see Section~ \ref{sec:datasetrepairthemall}).
The first three rows give the prediction results for the considered subsets of ODS features, i.e., 150 code description features, 26 repair pattern features, and 26 contextual features. The fourth to sixth rows give the prediction results with the K best features, where K is defined as 50, 100, and 150.
}

\TSErevision{
Those results allow us to deepen our understanding of ODS features as follows.
1) ODS with all features outperforms ODS with any subset of features: ODS achieves 93.66\% of accuracy, which is significantly higher the accuracy with only code description features (85.74\%), repair pattern features  (67.34\%) and contextual features  (89.34\%). This result suggests that the three groups of ODS features complement each other and capture separated characteristics of overfitting;
2) The relation between the number of selected features and the effectiveness of ODS is monotonous: more ODS features contribute to higher accuracy.  For example, the accuracy of the best 150 features is 92.44\%, which outperforms the accuracy with the best 100  features (91.32\%) and the best 50 features (83.69\%).  This confirms that the ODS features are not redundant, and all ODS features are positively contributed to the ODS effectiveness in identifying overfitting patches; 
3) With only a half number of ODS features, e.g. 100 features, ODS still can be considered effective. 
This means that in practice, when resources are limited to extract all features, practitioners could still achieve good accuracy by only considering the top-K subset of features.
}

\begin{figure}[t!]
 \centering
\includegraphics[width=0.45\textwidth]{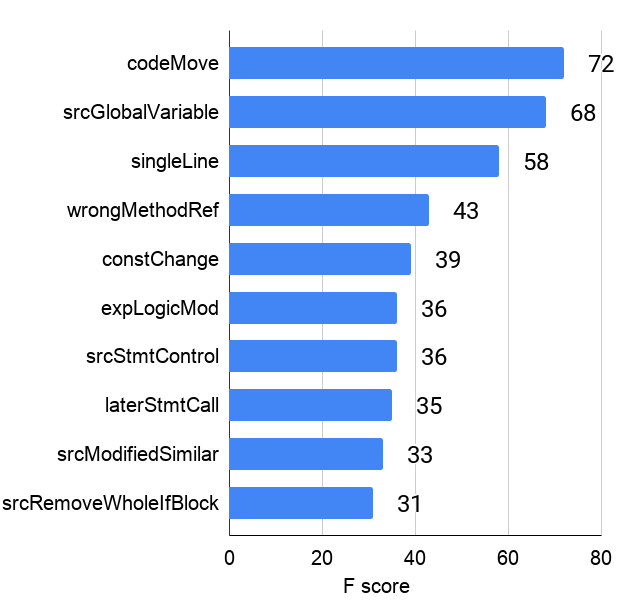}
\caption{The top-10 important ODS features.}
\label{fig-feature-important-rank}
\end{figure}

\begin{figure*}[t!]
 \centering
\includegraphics[width=0.95\textwidth,height=6cm]{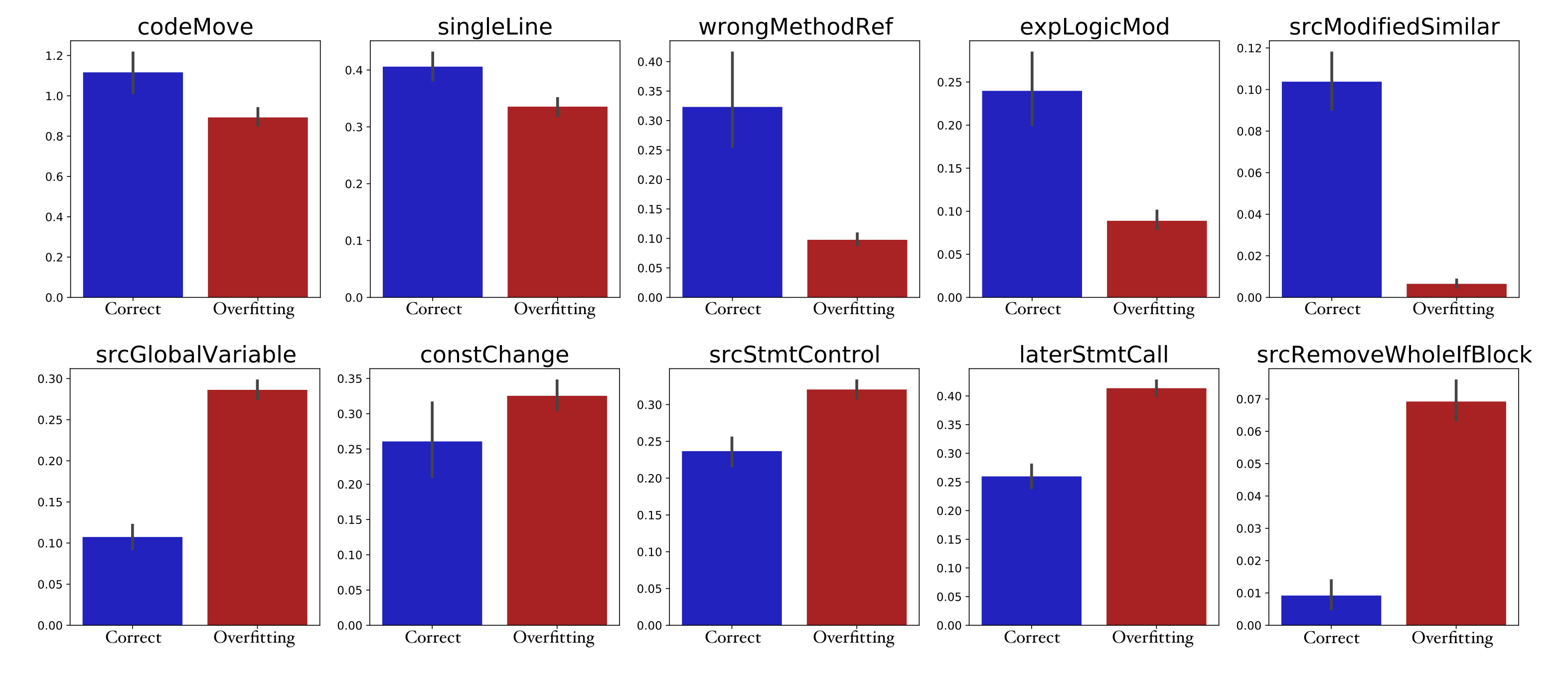}
\caption{Most important ODS features according to positive and negative correlation. The height of rectangles shows the mean value of features in a patch group and the cap of the bar shows the 95\% confidence interval. The \textcolor{blue}{blue} and \textcolor{red}{red} bars indicate the correct and overfitting patches, respectively.}
\label{fig-feature-study}
\end{figure*}

\TSErevision{Now we look at the important feature analysis. 
\autoref{fig-feature-important-rank} presents the  top-10   important ODS features that contribute to the accuracy of ODS. 
The X-axis gives the name of the features while the Y-axis shows the important values by their F-score, where F-score indicates the number of times a feature is used to split the data across all trees \cite{fscore}. 
The larger the F-score is, the more important the feature is.
For example,  feature \textit{codeMove} is shown as the most important ODS features.
Futhermore, to analyze how these features positively and negatively correlated to ODS accuracy, we present them in \autoref{fig-feature-study}.
}
In  \autoref{fig-feature-study},
the blue and red bars show the mean values of the feature in the set of correct and overfitting patches, respectively. 
The caps indicate the 95\% confidence interval of the mean based on boosting.
The 5 features presented in the top of the figure are negatively correlated: the lower the value, the more likely it is to be an overfitting patch.
The 5 features presented in the bottom of the figure  are positively correlated: the higher the value, the more likely it is to be an overfitting patch.

\emph{Negatively correlated features. }
The half top of \autoref{fig-feature-study} presents five ODS features that are negatively correlated to the overfitting patches: \textit{codeMove}, \textit{singleLine}, \textit{wrongMethodRef}, \textit{expLogicMod}, and \textit{srcModifiedSimilar}.
These can be interpreted as follows:
1) Correct patches tend to be lower granularity fix: feature \textit{singleLine} and \textit{expLogicMod} indicate the correct patches are more likely in a single line modification than overfitting patches and modifications are more likely to be made at the expression level.
2) Correct patches tend to be similar to existing code: feature \textit{codeMove}, \textit{wrongMethodRef} and \textit{srcModifiedSimilar} all mean modifications of existing code.
Feature \textit{codeMove} indicates a move of existing code to a different location, \textit{wrongMethodRef} means the fix requires an invocation of another existing method, and \textit{srcModifiedSimilar} indicates that the fix is similar to the existing code in the buggy program.
Notably, the significant difference of \textit{wrongMethodRef} between correct patches and overfitting patches suggests a limitation of program repair tools: they should consider more the strategy of replacing the wrong method invocations with existing methods. This also confirms the finding of \cite{emse18-repair-import-bugs} that automated repair techniques are less capable of fixing defects that require new function calls.
Also, the similarity of the modified code on overfitting patches is significantly lower than for correct patches (\textit{srcModifiedSimilar}).
This suggests that the program repair researchers can consider more the similarity of patch code to improve the effectiveness of program repair tools, as done in \cite{MechtaevICSE18-reference,capgen-ICSE18}.

\emph{Positively correlated features. }
The half bottom of \autoref{fig-feature-study} presents five ODS features that are positively correlated to the overfitting patches: \textit{srcGlobalVariables}, \textit{constChange}, \textit{srcStmtControl}, \textit{laterStmtCall} and \textit{srcRemoveWholeIfBlock}.
This can be interpreted as follows:
1) Overfitting patches tend to use more global variables (\textit{srcGlobalVariables}) and constant values (\textit{constChange}) than correct patches for a patch generation.
2) Overfitting patches are generated at a higher granularity: adding a control statement (\textit{srcStmtControl}) and removing a whole branch block (\textit{srcRemoveWholeIfBlock}) are high granularity transformations. This is consistent with the aforementioned finding that correct patches tend to be lower granularity fix based on negative correlation.
This high proportion of high granularity transformations come from program repair systems based on genetic programming (e.g., Arja and jGenProg): they often insert, update or delete the whole AST nodes corresponding to a faulty statement or block (e.g., the whole branch block) rather than focusing on expression level changes.

Beyond the features not discussed above, we also note that there are many other contextual features that are important according to correlation:
feature \textit{laterStmtCall} (shown in the fourth of the second row in \autoref{fig-feature-study}) shows that overfitting patches tend to have a method invocation after the patched location and feature \textit{typeofFaultyStatementParent} states that the type of the parent of faulty statement is important because it gives the scope information of a patched code (e.g., a parent type can be a class, method or block).

Contextual features provide important information and are even more valuable when considered in combination with the modification code features. 
For example, if the features of the patch transformations are similar to the features from the surrounded block, the patch tends to be overfitting. This similarity means that the patch code may duplicate its surrounding code, which is less likely to be correct. 
This confirms one of the core intuitions of ODS: the context is important, the correctness of a patch is not only related to the patched code but also to the code surrounding it. 
By considering the patched code and the context code together, ODS achieves a high accuracy for overfitting patch detection.

\begin{mdframed}
Answer to RQ4:
Our original feature analysis shows that all ODS features are useful, and contribute to the final effectiveness. 
A major advantage of ODS is that the features are explainable: one can look at features with high positive (or negative correlation) with overfitting and understand the overfitting diagnostic of ODS. 
\end{mdframed}

\section{Discussion}

\subsection{Threats to Validity}
\label{sec:threats}
We now discuss the threats to the validity of our results. 

\emph{Threats to internal validity.}
Bugs in the implementation of ODS could be a threat to the validity of our results.
To mitigate that threat, we have written unit tests to guarantee the correct behavior of the ODS tool.
We also manually inspected the results from ODS.
For some patches, we have manually compared the extracted code features and the source code to make sure they are consistent.
\TSErevision{Another internal threat relates to the re-implementation of Prophet in Java may contain potential defects.} To reduce those threats, we make all the source code and results publicly available for future researchers to check our code~\cite{ourrepo}.

\emph{Threats to external validity.} 
A threat to external validity relates to whether the performance of ODS
generalizes to arbitrary Java projects.
To mitigate this threat, 
we took a special in performing a large experiment:
1) on  \numprint{10302} patches written in Java;
2) from three Java bug benchmarks (Defects4J, Bugs.jar and Bears);
3) covering 95 open-source projects.
To our knowledge, this is the largest number of projects ever reported in this research field and this is arguably good for external validity.

\emph{Threats to construct validity.} 
The parameters involved in this study (shown in \autoref{sec:parameters}) may impact the reported results.
To mitigate this, we make publicly available the values of the hyperparameters we used (in particular, those we used on the XGBoost framework). 
Per our experiments, there is no evidence that the performance would dramatically change with other values of hyperparameters.

\begin{figure}[t!]
  \centering
\includegraphics[width=0.5\textwidth,height=6cm]{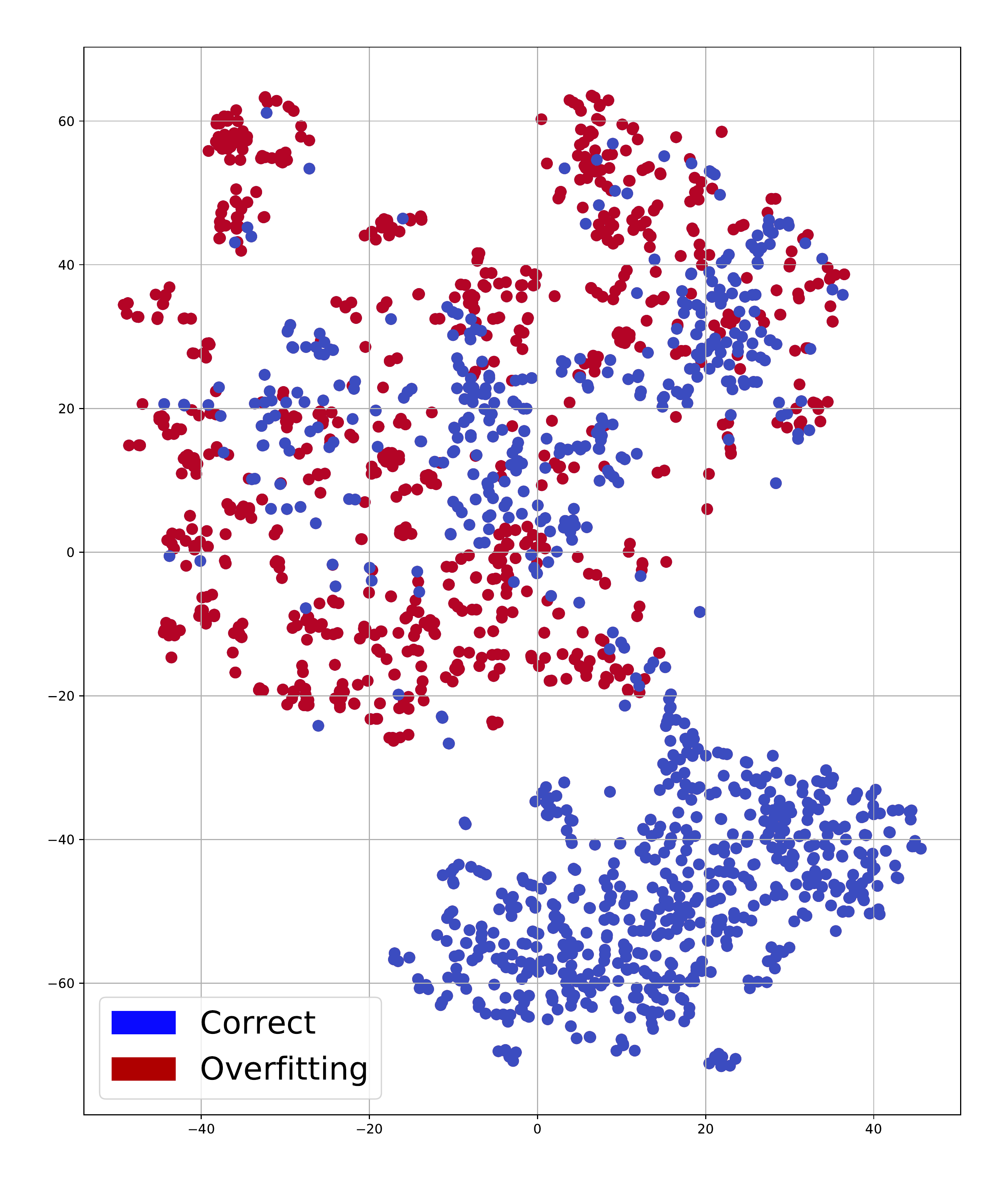}
\caption{t-SNE projection of a random sample of 1000 correct (blue) and 1000 overfitting (red) patches.}
\label{fig-features-distribution}
\end{figure}

\vspace{1cm}
\subsection{False Positives}

\TSErevision{
Recall that our RQ1 result reports that over \numprint{248} correct patches, \numprint{182} patches are rightly classified as correct and 66 patches are wrongly classified as overfitting (false positive). And our RQ3 reports that over \numprint{1725} correct patches, \numprint{1398} patches are rightly classified as correct and 327 patches are wrongly classified as overfitting. 
}

Now we investigate the reasons for which ODS yields false positives. 
First, we visualize the ODS features distribution. \autoref{fig-features-distribution} presents the distribution of correct and overfitting patches onto a two-dimensional map of ODS features with t-SNE clustering \cite{tsne}. The figure contains 2000 randomly selected patches, 1000 correct patches and 1000 overfitting patches, from the three considered benchmarks presented. 
The blue points represent the correct patches while the red points represent the overfitting patches.
As seen in \autoref{fig-features-distribution}, the majority of correct and overfitting patches form two distinct clusters: the overfitting patches cluster in the top while the correct patches cluster at the bottom. 
This shows that ODS indeed uses the code features to distinguish these two categories of patches.
Nevertheless, there exists a small overlap of overfitting patches and correct patches in the cluster of overfitting patches (top area).
This indicates that indeed some correct patches with similar code features as overfitting patches are wrongly classified as overfitting, and this is the cause of false positives. 
We manually investigate a sample of those false positive cases and we could identify two main reasons why that misclassification happens. 
First, some correct patches are complex and spread over different fixing locations and different buggy files (e.g., Chart-18 from Defects4J): this leads to noisy features and confuses the ODS classifier. 
Second, some patches are not perfectly described by the ODS features (as discussed in Section~\ref{sec:odsvspatchsim} and shown in the patch of \autoref{lst:fpmath70and73}). 
Finally, we note that there is no a single overfitting patch presents in the blue cluster of correct patches at the bottom: this suggests that in the most conservative configuration, ODS can achieve an accuracy of 100\%, i.e., ODS can deliver only correct program repair patches to developers and practitioners.


\section{Related work}


\label{sec:related-work}
\subsection{Overfitting Assessment without Additional Oracles} 
\label{subsec:related-work-1}

Researchers have proposed techniques for detecting overfitting patches without the use of additional oracles, for example, the human-written patches.
We now summarize the most important works on this line.

Long and Richard propose Prophet \cite{prophet} for ranking program repair patches in the C language.
ODS and Prophet share the same hypothesis that cross-project code features capture universal correctness properties for prioritizing APR patches. 
Beyond considering different programming languages, ODS and Prophet differ in their features and learning strategies.
W.r.t. features, Prophet proposes 30 atomic features and 5 modification features.  
The features of ODS are more comprehensive than Prophet's: they comprise repair patterns and contextual code features. 
W.r.t. learning strategies, ODS is trained with both overfitting and correct real-world patches generated by different APR systems, while Prophet only considers as input correct human-written patches. 
Along the same line, the overfitting training labels are not guaranteed in the Prophet experiment, while we found the ODS training dataset on a strict check on the plausibility of patches.

Tan et al. \cite{anti-pattern} propose Anti-patterns to identify whether code transformations in the patch fall into the category of pre-defined overfitting patterns. 
Anti-patterns and ODS are both static patch assessment techniques that encode human knowledge for overfitting patch detection.
On the evaluation of the 139 patches from the dataset of Xiong et al. \cite{patchsim}, 
Anti-patterns succeed in classifying 27 overfitting patches with one misclassification of a correct patch, while ODS identifies 70 overfitting patches. 
Beyond the higher effectiveness of ODS, we note that ODS and Anti-patterns are different in their strategies:   
Anti-patterns only analyze the transformations involved in a patch, while ODS considers the transformation structure, but in conjunction with other sources of information: the characteristics of fixing ingredients and the context information (i.e., the code around the transformations).

Opad  \cite{Opad} and Fix2Fit \cite{Gao:crashprepair} are two patch assessment techniques based on the creation of implicit oracles to detect overfitting patches that introduce crashes or memory-safety problems. 
We find the following difference between the ODS and these two techniques. 
First, Opad and Fix2Fit are dynamic patch assessment techniques, while ODS is a static  technique. 
Second, Opad and Fix2Fit target specific types of overfitting patches, those that introduce crashes (e.g., runtime exception), while ODS makes no assumption on the type of overfitting patches.

Tian et al. \cite{ase20Bert} employ BERT to automatically extract code features. 
ODS is significantly different from their work in terms of interpretability. Tian et al. and Csuvik et al. extract the code features with the BERT framework and consequently, those features black box and hard to interpret. On the contrary, the overfitting diagnostic of ODS are manual-crafted features and are interpretable.

Wang et al. \cite{ASE20Wang} leverage on eight static code features from three existing techniques (S3 \cite{s3}, SSFix \cite{ssFix}  and CapGen \cite{capgen-ICSE18}) for overfitting patch assessment.
As shown in our experiment, the ODS features results in both higher precision and recall compared to the eight code features considered in their work.


\subsection{Overfitting Assessment with Additional Oracles}

There is research on employing techniques to generate oracles based on the human-written patch considered as the ground truth. 

Xin and Reiss \cite{issta17-difftgen} propose DiffTGen to identify overfitting patches with tests generated by Evosuite \cite{evosuite} and oracles generated on human-written patches.
Those tests aim to detect behavioral differences between the program repair patch and the human-written patch.

Le et al. \cite{Le:overfitting, le:reliability-patch-assess} evaluate the overfitting problem in semantics-based techniques with automated test case generation techniques and ground truth patches. They further investigate the reliability of automatic patch correctness assessments with DiffTGen and random-based test generation technique Randoop \cite{randoop} by comparing them with manual assessments done by 35 professional developers \cite{le:reliability-patch-assess}.  
They found that it is not enough to only use these techniques (e.g., manual assessment and additional tests) to evaluate the effectiveness of program repair approaches.

Yu et al. \cite{zhongxing-EMSE18} and Ye et al. \cite{quixbugs} employ Evosuite \cite{evosuite} to automatically generate additional test cases based on the human-written patches on two benchmarks: Defects4J and QuixBugs, respectively. 
Ye et al. \cite{drr} present a large-scale experiment on using Evosuite \cite{evosuite} and Randoop \cite{randoop} to automatically generate test cases. They use those generated tests for validating the patch assessment manually done  previously by researchers. By doing so, they detected 12 mislabeled patches (i.e., overfitting patches classified as correct) from previous experiments, which shows the complexity and error-proneness of manual assessment. 

Yang and Yang \cite{ibf20-invariant} use the invariants generation to analyze the behavior of generated patches. 
Their study shows that the majority of plausible patches (92/96) expose different runtime invariants. 
Yet, the effectiveness of invariants for overfitting detection is not proven, the recent work of Wang et al. \cite{ASE20Wang} shows that this technique suffers from false positives to a large degree.

Wang et al. \cite{ASE20Wang} conduct an empirical study of oracle based patch assessment experiment to compare the effectiveness of existing dynamic techniques: DiffTGen, test generation based assessment, and invariant based patch assessment.
Their work shows that patch assessment techniques with additional oracles achieve significantly higher precision than patch assessment techniques without additional oracles. 

As mentioned, all those approaches need a ground truth patch to construct oracles which determine the correctness of patches.
On the contrary, ODS is able, as shown in this paper, to predict the correctness of patches without requiring a ground truth patch.

\subsection{Analyzing Programs with Features}
Different works have used features extracted from source code in the field of fault localization, program synthesis and automated program repair.

In the field of fault localization, works by \cite{xuan2017ranking, Kim:2019:PLF, Sohn:2017:FUC} have used dynamic and/or static features.
For instance, Kim et al. \cite{Kim:2019:PLF} present a learn-to-rank fault localization technique named PRINCE.
It first creates, based on 55 dynamic and static features, a ranking model for fault localization using genetic programming.
As a difference, beyond the different purposes, ODS is based exclusively on a larger different set of static features (202 features versus 55 features)
Similarly to PRINCE, Dam et al. \cite{Dam2018learning}
define an approach that automatically learns both semantic and syntactic features of code using the Long-Short-Term-Memory model with the goal of predicting vulnerabilities. Their evaluation showed it slightly outperforms PRINCE.

Yu et al. \cite{zhongxing-s4r} define a conditional random field that captures how certain repair transforms are applied to certain AST nodes, and then uses the learned model to predict transforms to be applied on buggy code. The focus is different, their work focuses on the synthesis of a patch transformation, while ODS focuses on overfitting patch detection.

Other works have focused on defect prediction based on code features. 
Their goal is to determine if a piece of code is buggy or clean.
For instance, Wang et al. \cite{wangsong-featurs} 
propose an approach that learns semantic features for defect prediction.
The approach takes tokens from the source code as input and generates semantic features from them, which are then used to build and evaluate the models for predicting defects.
Similarly, Shippey et al. \cite{SHIPPEY2019142} propose a defect prediction technique based on learning features from N-grams extracted from AST, and Hoang et al. \cite{msr19-deepjit} leverage the convolutional neural network (CNN)  to automatically encode the commit code and message for Just-In-Time (JIT) defect prediction. 
Our approach, on the contrary to those, uses carefully designed manual-crafty features that aim at capturing the syntactic and semantic of patch transformation.

\section{Conclusion}
We have presented ODS, a novel overfitting patch detection system that utilizes static code features.
To our knowledge, the considered set of features has ever been studied before in the context of overfitting patch assessment.
In our experiments, we have trained ODS to detect overfitting patches from \numprint{10302} program repair patches from three bug benchmarks (Defects4J, Bugs.jar and Bears), which makes it one of the largest learning-based experiments in this research field. 
The results of our evaluation show that ODS achieves an accuracy of 71.9\% in detecting overfitting patches from 26 projects, and outperforms the state-of-the-art. ODS is applicable in practice and can be employed as a post-processing procedure to classify the patches generated by different APR systems (e.g., \cite{coconut-issta20,sequencer}).
Notably, our experiment has evaluated whether patch assessment can be made in a project independent manner, which overfitting knowledge learned on one project and applied on another one.


Our future work will focus on dynamic features:  we plan to extract features from execution such as test diagnostics to understand the root cause of a bug \cite{bugRootCause-icse20}, and to combine them with the ODS static features for better overfitting patch classification.
Also, we are planning to adapt the ODS features to other tasks, such as patch ranking and bug clustering.

\section*{Acknowledgments}

This work was partially supported by the Wallenberg Artificial Intelligence, Autonomous Systems and Software Program (WASP) funded by Knut and Alice Wallenberg Foundation, by the Swedish Foundation for Strategic Research (SSF). Some experiments were performed on resources provided by the Swedish National Infrastructure for Computing.

\bibliography{references}
\bibliographystyle{plain}

\begin{IEEEbiography}
 [{\includegraphics[width=1in,height=1.25in,clip,keepaspectratio]{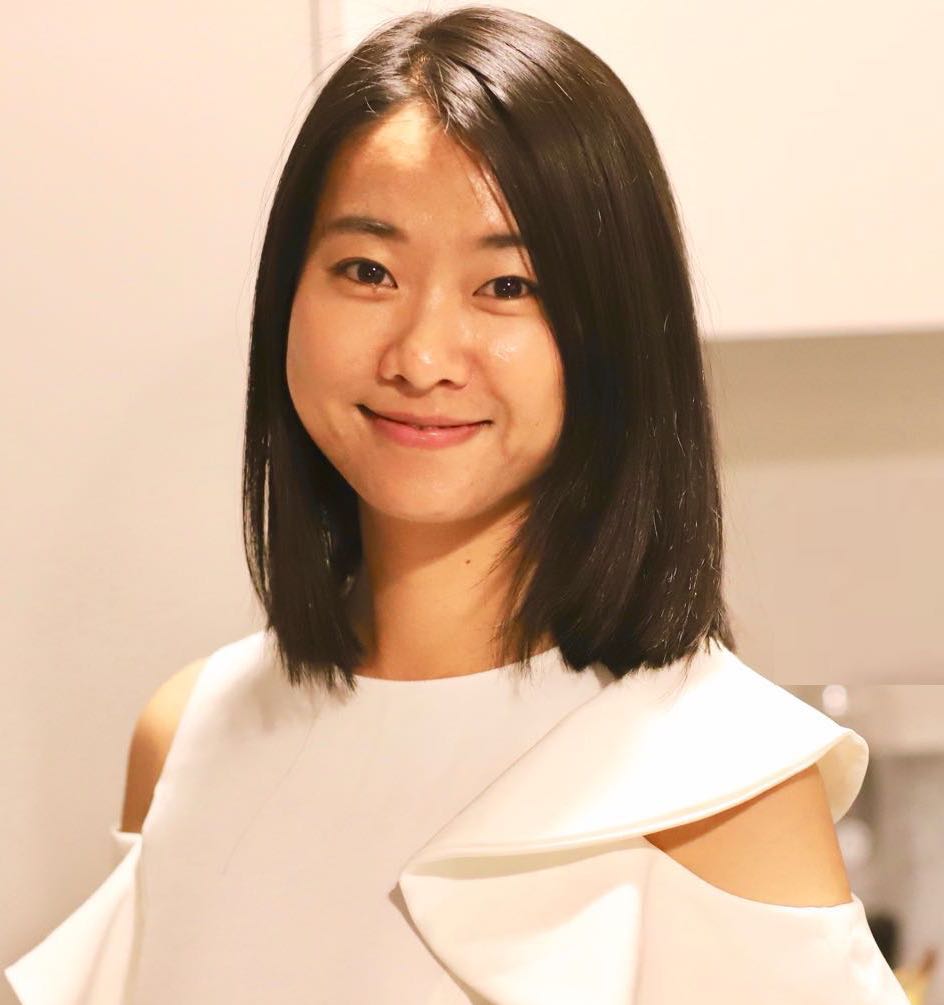}}]{He Ye}
is a PhD student in the department of theoretical computer science at KTH Royal Institute of Technology in Sweden. She received her BSc degree in software engineering from  Sichuan University and MSc degree in computer science from University of Tampere.  Her primary research interest lies in automatic program repair and software testing.
\end{IEEEbiography}

\vspace{-1cm}

\begin{IEEEbiography}
 [{\includegraphics[width=1in,height=1.25in,clip,keepaspectratio]{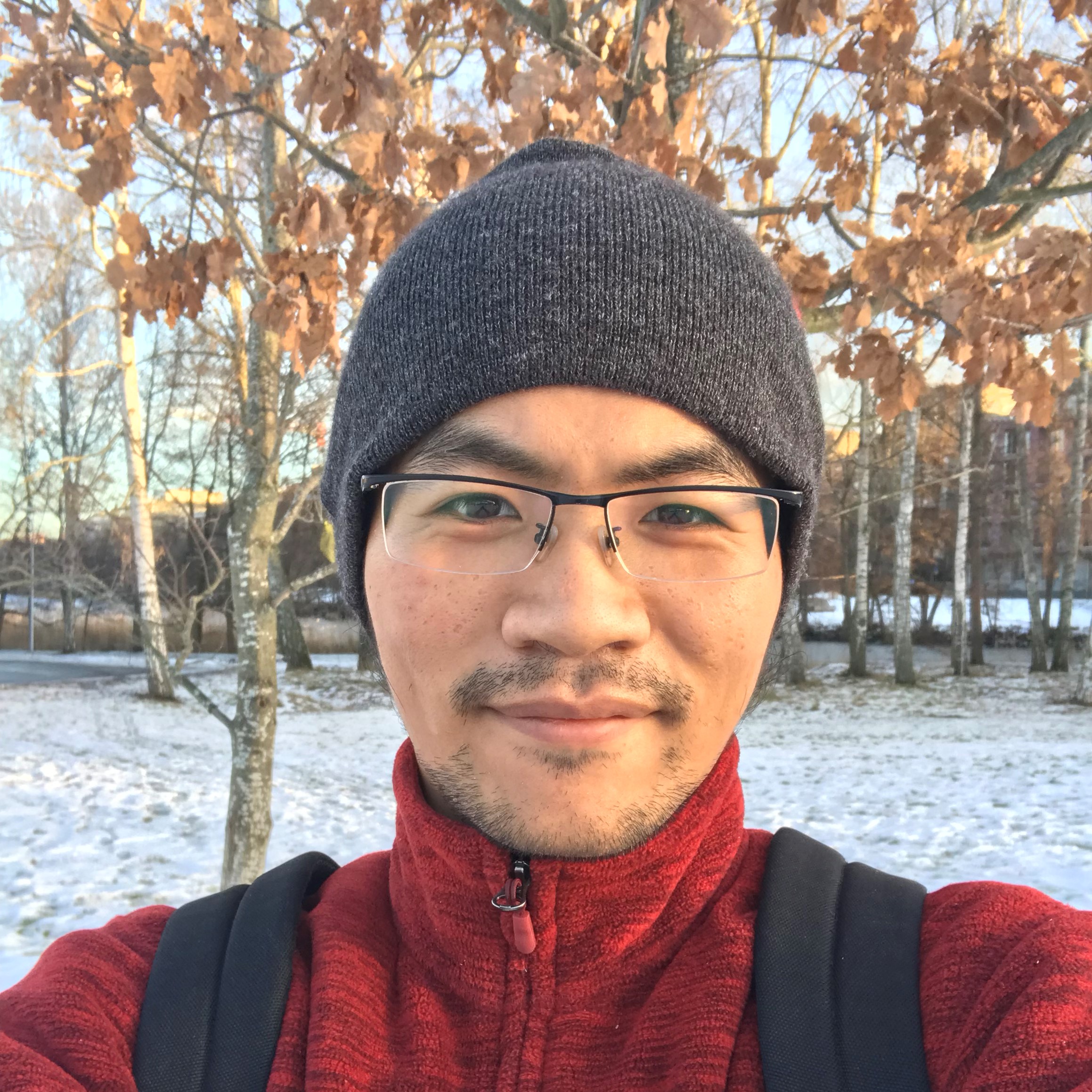}}]{Jian Gu}
is a Ph.D. student at the Software Evolution and Architecture Lab, University of Zurich. He obtained his BSc degree in computer science at Shandong University and MSc degree in machine learning at KTH Royal Institute of Technology. Jian is interested in machine learning solutions under the topic of software engineering.
\end{IEEEbiography}

\vspace{-1cm} 

\begin{IEEEbiography}[{\includegraphics[width=1in,height=1.25in,clip,keepaspectratio]{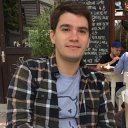}}]{Matias Martinez}
is an associate professor in the Université Polytechnique Hauts-de-France (France), and member of the LAMIH laboratory (UMR CNRS 8201). He got his PhD degree from University of Lille (France) and a Computer Science degree from UNICEN (Argentina).  
His research focuses on automated software engineering, software testing and programming languages.
\end{IEEEbiography}

\vspace{-1cm}

\begin{IEEEbiography}[{\includegraphics[width=1in,height=1.25in,clip,keepaspectratio]{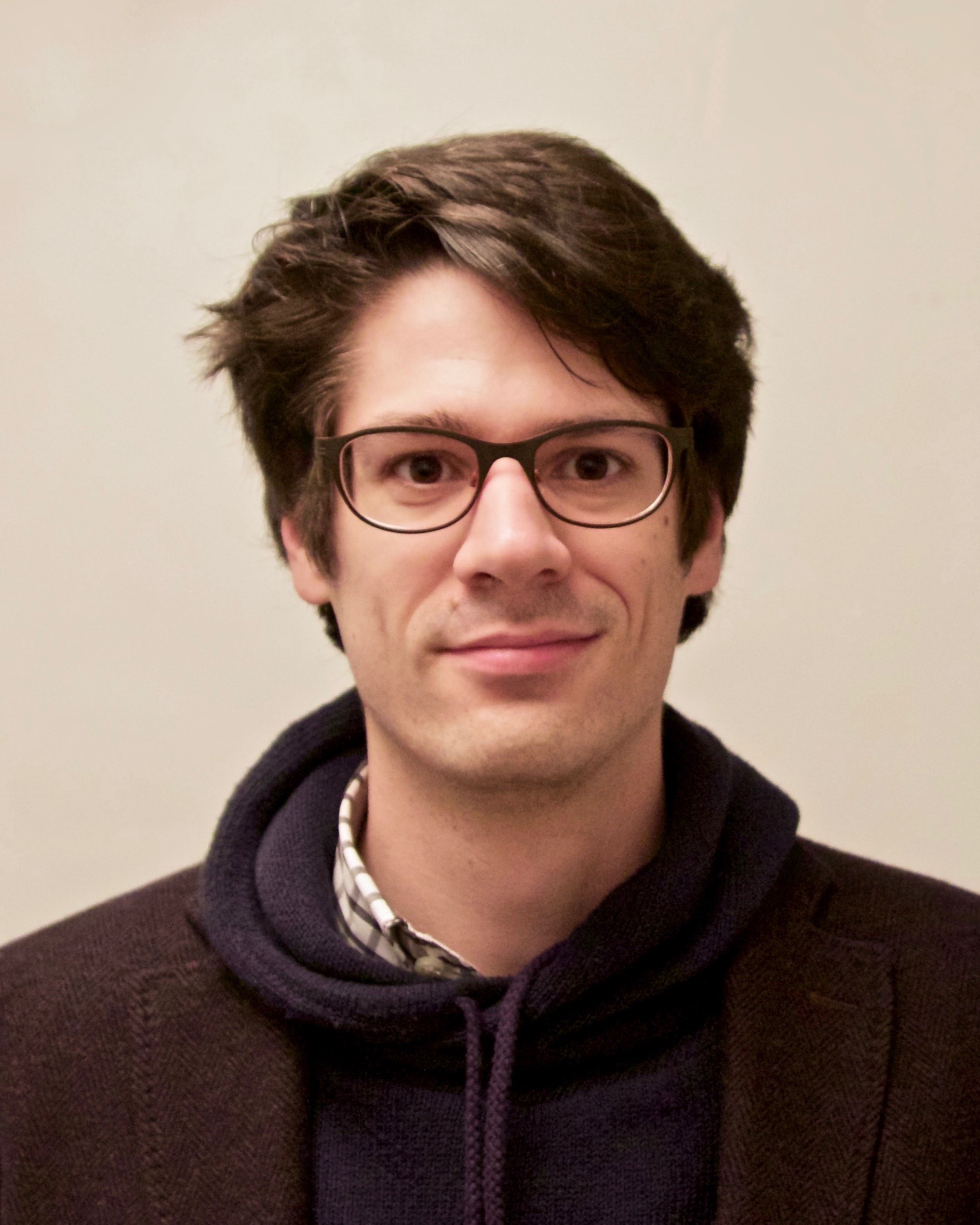}}]{Thomas Durieux}
is a Post-Doc at KTH Royal Institute of Technology. He obtained his PhD at University of Lille, France on automatic patch generation in production.  
His research interests are software debloating and automatic patch generation, bugs identification, developer interaction with bots. 
\end{IEEEbiography}

\vspace{-1cm}
\begin{IEEEbiography}[{\includegraphics[width=1in,height=1.25in,clip,keepaspectratio]{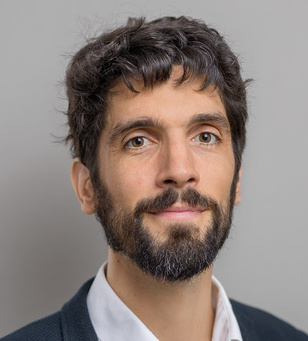}}]{Martin Monperrus}
is Professor of Software Technology at KTH Royal Institute of Technology. He was previously associate professor at the University of Lille and adjunct researcher at Inria. He received a Ph.D. from the University of Rennes, and a Master's degree from the Compiègne University of Technology. His research lies in the field of software engineering with a current focus on automatic program repair, program hardening and chaos engineering.

\end{IEEEbiography}

\end{document}